\documentclass[nocopyrightspace]{sigplanconf}
\usepackage{natbib}
\bibpunct();A{},
\let\cite=\citep
\usepackage{boxedminipage}     
\usepackage{proof}
\usepackage{amsmath} 
\usepackage{amsthm}
\usepackage{xspace}
\usepackage{paralist}

\newcommand{\tronly}[1]{#1}
\newcommand{\confonly}[1]{}

\usepackage{xspace}
\usepackage{graphicx}

\usepackage{amssymb}
\usepackage{amsfonts}
\usepackage{proof}
\usepackage{epsfig}
\usepackage{verbatim}

\renewcommand{\emptyset}{\varnothing}

\newcommand{\SB}[1]{[\![#1]\!]}

\newcommand{\eq}{\approx}

\newcommand{\hyp}[2][]{\infer[#1]{#2}{}}

\newcommand{\labelSec}[1]{\label{sec:#1}}
\newcommand{\refSec}[1]{Section~\ref{sec:#1}}
\newcommand{\labelFig}[1]{\label{fig:#1}}
\newcommand{\refFig}[1]{Figure~\ref{fig:#1}}

\newcommand{\labelThm}[1]{\label{thm:#1}}
\newcommand{\refThm}[1]{Theorem~\ref{thm:#1}}
\newcommand{\labelLem}[1]{\label{lem:#1}}
\newcommand{\refLem}[1]{Lemma~\ref{lem:#1}}

\newcommand{\To}{\Rightarrow}
\newcommand{\eval}{\Downarrow}

\newcommand{\tri}{\triangleright}

\newcommand{\nd}{\vdash}

\newtheorem{theorem}{Theorem}
\newtheorem{lemma}{Lemma}

\newcommand{\evalu}[4]{#1;#2 \vdash #3 \Rightarrow^\mathtt{U} #4}
\renewcommand{\eval}[3]{#1 \vdash #2 \Rightarrow #3}
\newcommand{\kas}{\kw{as}}

\newcommand{\subty}{\mathrel{{<}{:}}}

\newcommand{\XQueryBang}{XQuery!\xspace}

\newcommand{\muXQ}{$\mu$XQ\xspace}
\newcommand{\Lux}{\textsc{Lux}\xspace}
\newcommand{\Flux}{\textsc{Flux}\xspace}

\newcommand{\tylab}[3]{#1 :: #2 \To #3}
\newcommand{\wfin}[5]{#1 \vdash \bar{#2}~\kin~#3 \to #4 : #5}
\newcommand{\wfineff}[6]{#1 \vdash \bar{#2}~\kin~#3 \to #4 : #5 \mathrel{\&} #6}

\newcommand{\wfiter}[4]{#1 \vdash_{\kiter} \{#2\}~ #3~ \{#4\}}

\newcommand{\wfupd}[5]{#1 \vdash^{#2} \{#3\}~#4~\{#5\}}

\newcommand{\wf}[3]{#1 \vdash #2:#3}
\newcommand{\wfeff}[4]{#1 \vdash #2:#3 \mathrel{\&} #4}

\newcommand{\wfupdeff}[6]{#1 \vdash^#2 \{#3\}~ #4~ \{#5\}\mathrel{\&} #6}
\newcommand{\wfitereff}[5]{#1 \vdash_{\kiter} \{#2\}~ #3~ \{#4\}\mathrel{\&} #5}

\newcommand{\Atoms}{\mathit{Atom}}
\newcommand{\Stmt}{\mathit{Stmt}}

\newcommand{\Lab}{\mathit{Lab}}
\newcommand{\Test}{\mathit{Test}}
\newcommand{\Expr}{\mathit{Expr}}
\newcommand{\Dir}{\mathit{Dir}}
\newcommand{\Types}{\mathit{Type}}
\newcommand{\Tree}{\mathit{Tree}}
\newcommand{\Bool}{\mathit{Bool}}
\newcommand{\Val}{\mathit{Val}}

\newcommand{\Var}{\mathit{Var}}
\newcommand{\TVar}{\mathit{TVar}}
\newcommand{\TyVar}{\mathit{TyVar}}

\newcommand{\emptyseq}{\texttt{()}}
\newcommand{\kw}[1]{\mathtt{#1}}
\newcommand{\ktransform}{\kw{transform}}
\newcommand{\kby}{\kw{by}}
\newcommand{\transform}[2]{\ktransform~#1~\kby~#2}

\newcommand{\kchildren}{\kw{children}}
\newcommand{\kchild}{\kw{child}}

\newcommand{\ksnapshot}{\kw{snapshot}}
\newcommand{\snapshot}[2]{\ksnapshot~#1~\kin~#2}
\newcommand{\kbool}{\kw{bool}}
\newcommand{\kstring}{\kw{string}}
\newcommand{\kreturn}{\kw{return}}
\newcommand{\ktrue}{\kw{true}}
\newcommand{\kfalse}{\kw{false}}

\newcommand{\kleft}{\kw{left}}
\newcommand{\kright}{\kw{right}}

\newcommand{\kinsert}{\kw{insert}}

\newcommand{\kdelete}{\kw{delete}}
\newcommand{\krename}{\kw{rename}}
\newcommand{\kif}{\kw{if}}
\newcommand{\kthen}{\kw{then}}
\newcommand{\kelse}{\kw{else}}
\newcommand{\klet}{\kw{let}}
\newcommand{\kin}{\kw{in}}

\newcommand{\kskip}{\kw{skip}}

\newcommand{\kfor}{\kw{for}}

\newcommand{\kiter}{\kw{iter}}

\newcommand{\kTO}{\kw{TO}}
\newcommand{\kINTO}{\kw{INTO}}
\newcommand{\kIN}{\kw{IN}}
\newcommand{\kBY}{\kw{BY}}
\newcommand{\kFROM}{\kw{FROM}}
\newcommand{\kVALUE}{\kw{VALUE}}
\newcommand{\kINSERT}{\kw{INSERT}}
\newcommand{\kDELETE}{\kw{DELETE}}
\newcommand{\kRENAME}{\kw{RENAME}}
\newcommand{\kREPLACE}{\kw{REPLACE}}
\newcommand{\kUPDATE}{\kw{UPDATE}}
\newcommand{\kWITH}{\kw{WITH}}
\newcommand{\kIF}{\kw{IF}}
\newcommand{\kTHEN}{\kw{THEN}}

\newcommand{\kLET}{\kw{LET}}
\newcommand{\kWHERE}{\kw{WHERE}}
\newcommand{\kFIRST}{\kw{LAST}}
\newcommand{\kLAST}{\kw{FIRST}}
\newcommand{\kAS}{\kw{AS}}
\newcommand{\kBEFORE}{\kw{BEFORE}}
\newcommand{\kAFTER}{\kw{AFTER}}
\newcommand{\kbefore}{\kw{BEFORE}}
\newcommand{\kafter}{\kw{AFTER}}
\newcommand{\kfirst}{\kw{FIRST}}
\newcommand{\klast}{\kw{LAST}}
\newcommand{\ifthenelse}[3]{\kif~#1~\kthen~#2~\kelse~#3}

\newcommand{\IFTHEN}[2]{\kIF~#1~\kTHEN~#2}
\newcommand{\letin}[2]{\klet~ #1~\kin~#2}
\newcommand{\LETIN}[2]{\kLET~ #1~\kIN~#2}
\newcommand{\forreturn}[2]{\kfor~#1~\kreturn~#2}

\newcommand{\elab}[2][]{\textsf{\textbf{[}}#2\textsf{\textbf{]}}_{#1}}
\newcommand{\norm}[2][]{\textsf{\textbf{[}}#2\textsf{\textbf{]}}_{#1}}
\newcommand{\ichildren}{\mathit{children}}

\newcommand{\delete}[1]{ \kDELETE~#1}
\newcommand{\deletefrom}[1]{ \kDELETE~\kFROM~#1}
\newcommand{\insertvalue}[3]{ \kINSERT~#1~#2~\kVALUE~#3}
\newcommand{\insertinto}[3]{ \kINSERT~\kAS~#1~\kINTO~#2~\kVALUE~#3}
\newcommand{\rename}[2]{\kRENAME~#1~\kTO~#2}
\newcommand{\replace}[2]{\kREPLACE~#1~\kWITH~#2}
\newcommand{\update}[2]{\kUPDATE~#1~\kBY~#2}
\newcommand{\replacein}[2]{\kREPLACE~\kIN~#1~\kWITH~#2}
\newcommand{\Upd}{\mathit{Upd}}
\newcommand{\Path}{\mathit{Path}}

\newcommand{\knode}{\mathtt{node}}
\newcommand{\ktext}{\mathtt{text}}

\newcommand{\npath}[1]{\norm[\Path]{#1}}
\newcommand{\nupd}[1]{\norm[\Upd]{#1}}
\newcommand{\nstmt}[1]{\norm[\Stmt]{#1}}
\newcommand{\wfupdstmt}[4]{#1 \vdash_\Stmt \{#2\}~#3~\{#4\}}
\newcommand{\wfupdsimp}[4]{#1 \vdash_\Upd \{#2\}~#3~\{#4\}}
\newcommand{\wfupdpath}[5]{#1 \vdash_\Path \{#2\}~#3 \leadsto (#4,#5)}
\newcommand{\wfupdpaths}[4]{\vdash_\Path \{#1\}~#2 \leadsto (#3,#4)}
\newcommand{\wfupds}[3]{\vdash^{1} \{#1\}~#2~\{#3\}}
\newcommand{\wfupdstmts}[3]{\vdash_\Stmt \{#1\}~#2~\{#3\}}
\newcommand{\wfs}[3]{\vdash \{#1\}~#2:#3}
\newcommand{\wfupdfilt}[5]{#1 \vdash #2 :: \phi \leadsto (#4, #5)}
\newcommand{\subst}[1]{\langle #1 \rangle}

\newcommand{\procdecl}[3]{#1:#2\To#3}

\newcommand{\typedecl}[2]{\mathtt{type}~#1=#2}

\theoremstyle{definition}
\newtheorem{definition}{Definition}

\newcommand{\maybe}[1]{#1{?\!\!?}}
\newcommand{\defeq}{\triangleq}
\newcommand{\wfdecl}[1]{\vdash_{\mathit{Decl}} #1}

\begin{document}

\confonly{\conferenceinfo{ICFP'08,} {September 22--24, 2008, Victoria, BC, Canada.}}
\confonly{\CopyrightYear{2008}}
\confonly{\copyrightdata{978-1-59593-919-7/08/09}}

\confonly{\title{\Flux: FunctionaL Updates for XML}}
\tronly{\title{\Flux: FunctionaL Updates for XML (extended report)}}
\authorinfo{James Cheney}
           {University of Edinburgh}
           {jcheney@inf.ed.ac.uk}
\maketitle
\begin{abstract}
  XML database query languages have been studied extensively, but XML
  database updates have received relatively little attention, and pose
  many challenges to language design.  We are developing an XML update
  language called \Flux, which stands for FunctionaL Updates for XML,
  drawing upon ideas from functional programming languages.  In prior
  work, we have introduced a core language for \Flux with a clear
  operational semantics and a sound, decidable static type system
  based on regular expression types.
  
  Our initial proposal had several limitations. First, it lacked
  support for recursive types or update procedures.  Second, although
  a high-level source language can easily be translated to the core
  language, it is difficult to propagate meaningful type errors from
  the core language back to the source.  Third, certain updates are
  well-formed yet contain \emph{path errors}, or ``dead''
  subexpressions which never do any useful work.  It would be useful
  to detect path errors, since they often represent errors or
  optimization opportunities.

  In this paper, we address all three limitations.  Specifically, we
  present an improved, sound type system that handles recursion.  We
  also formalize a source update language and give a translation to
  the core language that \emph{preserves} and \emph{reflects}
  typability.  We also develop a \emph{path-error analysis} (a form of
  dead-code analysis) for updates.
\end{abstract}

\section{Introduction}

XQuery is a World Wide Web Consortium (W3C) standard, typed, purely
functional language intended as a high-level interface for querying
XML databases.  It is meant to play a role for XML databases analogous
to that played by SQL for relational databases.  The operational
semantics and type system of XQuery 1.0 has been formalized
\cite{xquery-semantics-w3c-20070123}, and the W3C recently endorsed the
formal semantics as a \emph{recommendation}, the most mature phase for
W3C standards.

Almost all useful databases change over time.  The SQL standard
describes a \emph{data manipulation language} (DML), or, more briefly,
\emph{update language}, which facilitates the most common patterns of
changes to relational databases: insertion, deletion, and in-place
modification of rows, as well as addition or deletion of columns or
tables.  Despite the effectful nature of these operations, their
semantics is still relatively clear and high-level.  SQL updates are
relatively inexpressive, but they are considered sufficient for most
situations, as witnessed by the fact that in many SQL databases, data
can \emph{only} be updated using SQL updates in transactions.
Moreover, the presence of SQL updates does no damage to the purely
functional nature of SQL queries: updates are syntactically distinct
from queries, and the language design and transactional mechanisms
ensure that aliasing difficulties cannot arise, even when an update
changes the structure of the database (for example, if a column is
added or removed from a table).

The XQuery standard lacks update language features analogous to SQL's
DML.  While XML querying has been the subject of a massive research
and development effort, high-level XML update languages have received
comparatively little attention.  Many programming languages for
transforming \emph{immutable} XML trees have been studied, including
XML stylesheets (XSLT \cite{clark99xslt}), and XML programming
languages such as XDuce, CDuce, Xtatic, or OCamlDuce
\cite{hosoya03toit,benzaken03icfp,DBLP:conf/planX/GapeyevGP06,frisch06icfp}.
However, these languages are not well-suited to specifying updates.
Updates typically change a small part of the document and leave most
of the data fixed.  To simulate this behavior by transforming
immutable XML values one must explicitly describe how the
transformation preserves unchanged parts of the the input.  Such
transformations are typically executed by building a new version of
the document and then replacing the old one.  This is inefficient when
most of the data is unchanged.  Worse, XML databases may employ
auxiliary data structures (such as indices) or invariants (such as
validity or key constraints) which need to be maintained when an
update occurs, and updating a database by deleting its old version and
loading a new version forces indices and constraints to be
re-evaluated for the whole database, rather than incrementally.

Instead, therefore, several languages specifically tailored for
updating XML data \emph{in-place} have been proposed.  While the
earliest proposal, called XUpdate \cite{laux00xmldb}, was relatively
simple and has been widely adopted, it lacks first-class conditional
and looping constructs.  These features have been incorporated into
more recent proposals~\cite{DBLP:conf/sigmod/TatarinovIHW01,sur04planx,DBLP:conf/edbtw/GhelliRS06,chamberlin06ximep,ghelli07dbpl}.
The W3C is also developing a standard XQuery Update
Facility~\cite{xquery-update-w3c-10-20080314}.

Although they have some advantages over XUpdate, we argue that these
approaches all have significant drawbacks, because they unwisely
combine imperative update operations with XQuery's purely-functional
query expressions.  We shall focus our critique on XQuery!, since it
is representative of several other proposals, including the W3C's XQuery
Update Facility.

A defining principle of XQuery!  is that update operations should be
``fully compositional'', which \citet{DBLP:conf/edbtw/GhelliRS06} take
to mean that an update operation should be allowed anywhere in an
XQuery expression.  Thus, the atomic update operations such as
insertion, deletion, replacement, renaming, etc.  may all appear
within XQuery's query expressions.  Node-identifiers can be used as
mutable references.  To avoid nondeterminism, \XQueryBang fixes a
left-to-right evaluation order and employs a two-phase semantics that
first collects updates into a \emph{pending update list} by evaluating
an expression without altering the data, and then performs all of the
updates at once.  An additional operator called \verb|snap| provides
programmer control over when to apply pending updates.

\XQueryBang seems to sacrifice most of the good properties of XQuery.
Most equational laws for XQuery expressions are invalid for
\XQueryBang, and the semantics is highly sensitive to arbitrary
choices.  For example, consider the following \XQueryBang update.
\begin{verbatim}
for $x in $doc//a,
    $y in $doc//b
return (insert $y into $x, delete $x//c)
\end{verbatim}
Its behavior on two trees is shown in \refFig{xquerybang}.  In the
first example, consider input tree (a) with regular structure.
Running the above update on this tree yields the update sequence:
\begin{verbatim}
insert(1,<b><d/><b>); delete(4); 
insert(1,<b><e/><b>); delete(4);
insert(2,<b><d/><b>); delete(6); 
insert(2,<b><e/><b>); delete(6);
\end{verbatim}
The numbers refer to the node identifiers shown in
\refFig{xquerybang}(a) as superscripts.  When these updates are
performed, the result is output (b).  Note that each subtree labeled
$a$ in the output contains three $b$-subtrees, one corresponding to
the original $b$ and one for each occurrence of $b$ in the tree.  In
the second example, tree (c) is transformed to (d) via updates
\begin{verbatim}
insert(1,<b><c/><b>); delete(3); 
insert(1,<b><a><c/><a/><b/>); delete(3);
insert(5,<b><c/><b>); delete(6);
insert(5,<b><a><c/><a/><b/>); delete(6);
\end{verbatim}
Observe that both occurrences of $a$ have as subtrees both
occurrences of $b$ in the input.  This is because the snapshot
semantics of \XQueryBang first collects the updates to be performed,
then performs them in sequence.  Inserts always copy data from the
original version of the document, whereas deletes mark nodes in the
new version of the document for deletion.  This is why some
occurrences of $c$ remain below occurrences of $a$.  Although this is
not an update that a user would typically write, an implementation,
type system, or static analysis must handle all of the expressions in
the language, not just the well-behaved ones.

\begin{figure}
  \begin{center}
    \includegraphics[scale=0.6]{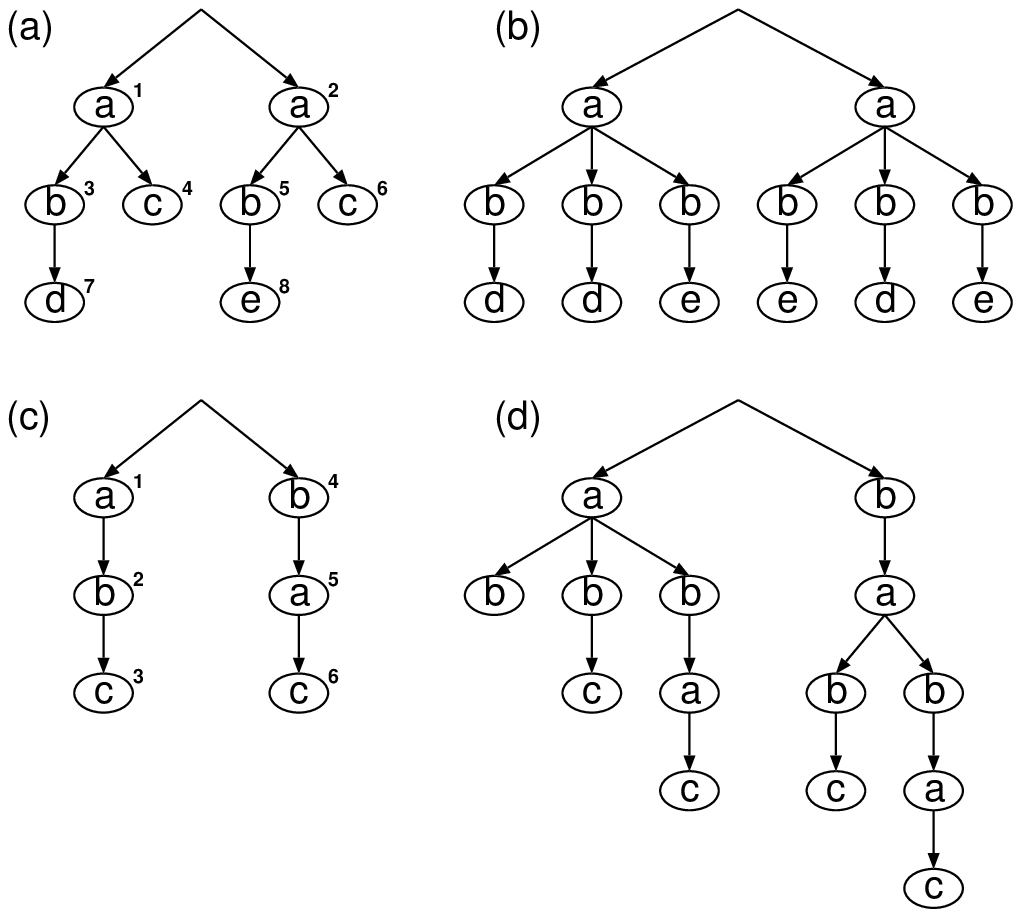}
  \end{center}
  \caption{\XQueryBang examples}\labelFig{xquerybang}
\end{figure}

Furthermore, the \XQueryBang approach seems quite difficult to
statically typecheck.  There are several reasons for this.  First, as
the examples in \refFig{xquerybang} show, the structure of the result
can depend on the data in ways difficult to predict using types.
Second, \XQueryBang also permits side-effects to be made visible
before the end of an update, using a ``snapshot'' operator called
\verb|snap|.  The \verb|snap| operator forces all of the delayed
side-effects of an expression to be performed.  This means that the
values of variables may change during an update, so it would be
necessary to update the types of variables to typecheck such updates.
Since variables may alias parts of the document, this requires a
nontrivial alias analysis.

We argue that the combination of features considered in \XQueryBang
and similar proposals are unnecessarily complex for the problem of
updating XML databases.  While high expressiveness is certainly a
reasonable design goal, we believe that for XML database updates,
expressiveness must be balanced against other concerns, such as
semantic transparency and static typechecking.  We believe that it is
worthwhile to consider an alternative approach that sacrifices
expressiveness for semantic clarity and the ability to typecheck and
analyze \emph{typical} updates \emph{easily}.

In previous work~\cite{DBLP:conf/planX/Cheney07}, we introduced a core
\emph{FunctionaL Update language for XML}, called
\Flux.\footnote{originally \Lux, for ``Lightweight Updates for XML''.}
\Flux is \emph{functional} in the same sense that imperative
programming in Haskell using monads\footnote{Technically, \Flux's
  approach to typechecking updates is closer to
  \emph{arrows}~\cite{hughes00scp}; however, we will not investigate
  this relationship in detail here.}  is functional.  Side-effects may
be present, but they are encapsulated using syntactic and type
constraints, so that queries remain purely functional.  \Flux provides
sufficient expressive power to handle most common examples of XML
database-style updates (e.g. all of the relational use cases in the
XQuery Update Facility Requirements
\cite{xquery-update-requirements-w3c-062006}), while avoiding
complications resulting from the interaction of unrestricted iteration
and unconstrained side-effects.

\Flux admits a relatively simple, one-pass operational semantics, so
side-effects can be performed eagerly; it can also be typechecked
using regular expression types, extending previous work on
typechecking
XQuery~\cite{hosoya05toplas,colazzo06jfp,xquery-semantics-w3c-20070123}.
The decidability of typechecking for the core language (with recursive
types and functions) was later established by \citet{cheney08esop}
along with a related result for an XQuery core language.
However, our preliminary proposal ~\cite{DBLP:conf/planX/Cheney07} had
several limitations.  This paper presents an improved design.  In
particular, our contributions relative to prior work are:
\begin{compactitem}
\item We extend the core language with recursive types and update
  procedures and provide a sound type system.
\item We adapt the idea of \emph{path-error analysis}, introduced by
  \citet{colazzo06jfp} for a core XML query language, to the setting
  of updates, and design a correct static analysis for conservatively
  under-approximating update path-errors.
\item We formalize a high-level \Flux source language and show how to
  translate to core \Flux.  We also present a source-level type system
  and prove that a source update is well-formed \emph{if and only if}
  its translation is well-formed.
\end{compactitem}

The structure of the rest of this paper is as follows.
\refSec{examples} briefly recapitulates the \Flux source language
introduced in \cite{DBLP:conf/planX/Cheney07} with a few examples.
\refSec{formalization} formalizes the core language and its
operational semantics; its type system is presented and proved sound
in \refSec{types}. \refSec{normalization} presents the translation
from the high-level \Flux language to Core \Flux and shows how to
typecheck high-level updates. \refSec{path-analysis} presents the
path-error analysis.  
We provide a more detailed comparison with related approaches in
\refSec{related}; \refSec{extensions} presents extensions and future
work; and \refSec{concl} concludes.

\confonly{Certain definitions and proofs have been omitted from this paper due
to space limitations, but can be found in the companion technical
report~\cite{flux-tr}.}
\tronly{Certain definitions and proofs have been placed in appendices.}

\section{Overview and examples}\labelSec{examples}
\subsection{Syntax}

As with many similar languages, particularly SQL,
XQuery~\cite{xquery-semantics-w3c-20070123}, and
CPL+~\cite{DBLP:conf/ssdbm/LiefkeD99}, we will introduce a high-level, readable
source language syntax which we will translate to a much simpler core
language.  We will later formalize the operational semantics and type
system for the core language.  In what follows, we assume familiarity
with XQuery and XPath syntax, and with XDuce-style regular expression
types.

The high-level syntax of \Flux updates is shown in
\refFig{lux-concrete}.  XQuery variables $\Var$ are typically written
\verb|$x|, \verb|$y|, etc. We omit the syntactic class $\Expr$
consisting of ordinary XQuery expressions, respectively.  Statements
$\Stmt$ include conditionals, let-binding, sequential composition, and
update statements $\Upd$, which may be guarded by a $\kWHERE$-clause.
We use braces to parenthesize statements $\{\Stmt\}$.  Updates $\Upd$
come in two flavors, \emph{singular} and \emph{plural}.  Singular
updates expect a single tree and are executed once for each selected
tree; plural updates operate on arbitrary sequences and are executed
on the children of each selected tree.  Singular insertions
(\verb|INSERT BEFORE/AFTER|) insert a value before or after each node
selected by the path expression, while plural insertions
(\verb|INSERT AS FIRST/LAST INTO|) insert a value at the beginning or
end of the child-list of each selected node.  Similarly, singular
deletes (\verb|DELETE|) delete individual nodes selected by the given
$\Path$, whereas plural deletes (\verb|DELETE| \verb|FROM|) delete the
child-list of each selected node.  Singular replacement
\verb|REPLACE WITH| replaces a subtree, while plural replacement
\verb|REPLACE| \verb|IN| replaces the content of a path with new
content.  The renaming operation \verb$RENAME TO$ is always singular;
it renames a subtree's label.  The $\kUPDATE~\Path~\kBY~\Stmt$
operation is singular; it applies $\Stmt$ to each tree matching
$\Path$.  Update procedure declarations are not shown but can be added
easily to the source language.

\begin{figure}[tb]
\[\begin{array}{rcl}
\Stmt &::=& \Upd~[\kWHERE~\Expr] \\
&|& \IFTHEN{\Expr}{\Stmt}\\
&|& \Stmt;\Stmt'\\
&|& \LETIN{\Var:=\Expr}{\Stmt}\\
&|& \{\Stmt\}\\
\Upd&::=& \insertvalue{(\kBEFORE|\kAFTER)}{\Path}{\Expr}\\
&|& \insertinto{(\kFIRST|\kLAST)}{\Path}{\Expr}\\
&|& \delete{[\kFROM]~\Path}\\
&|& \rename{\Path}{\Lab}\\
&|& \replace{[\kIN]~\Path}{\Expr}\\
&|& \update{\Path}{\Stmt}\\
\Path&::=& .  \mid \Lab \mid \knode() \mid \ktext() \\
&\mid& \Path/\Path         \mid \Var~\kAS~\Path \mid \Path[\Expr]
\end{array}\]
  \caption{Concrete syntax of \Flux updates.}\labelFig{lux-concrete}
\end{figure}

The \emph{path expressions} $\Path$ in \Flux are based on the XPath
expressions that are allowed in XQuery.  Paths include the empty path
$.$, sequential composition $\Path/\Path$, the XPath child axis tests
(\verb$text()$, \verb$\Lab$ and \verb$node()$), filters
($\Path[\Expr]$), and variable binding steps ($\Var~ \kAS~ \Path$).
The ``as'' path expression \verb|$x AS Path| (not present in XPath)
binds the subtree matching \verb|Path| to \verb|$x| in each iteration
of a path update.  We often write $*$ instead of $node()$.  We only
describe the syntax of paths used to perform \emph{updates}; arbitrary
XPath or XQuery expressions may be used in subqueries $\Expr$.

Both the \Flux source language described here and the core language
introduced later are case-insensitive with respect to keywords (like
XQuery); however, we use uppercase for the source language and
lowercase for the core language to prevent confusion.

\subsection{Execution model, informally}

In general, an update is evaluated as follows: The path expression is
evaluated, yielding a \emph{focus selection}, or a set of parts of the
updatable store on which the update \emph{focuses}.  The
\verb|WHERE|-clause, if present, is evaluated with respect to the
variables bound in the path and if the result is \verb|true| then the
corresponding basic update operation (insert, delete, etc.)  is
applied to each element of this set in turn.  Order of evaluation is
unspecified and the semantics is consistent with parallel evaluation
of iterations.

Unlike most other proposals, in \Flux, arbitrary XPath or XQuery
expressions cannot be used to select foci.  If this were allowed, it
would be easy to construct examples for which the result of an update
depends on the order in which the focus selection is processed.  For
example, suppose the document is of the form \verb|<a><b/></a>|.  If
the following update were allowed:
\begin{verbatim}
UPDATE //* BY { DELETE a/b;   RENAME * TO c }
\end{verbatim}
then the result would depend on the order in which the updates are
applied.  Two possible results are \verb|<c/>| and
\verb|<c><c/></c>|.  This nondeterministic behavior is difficult to
typecheck.  For this reason, we place severe restrictions on the path
expressions that may be used to select foci.

We identify two key properties which help to ensure that updates are
deterministic and can be typechecked.  First, \emph{an update can only
  modify data at or beneath its current focus}.  We call this
the \emph{side-effect isolation property}.  For example, navigating to
the focused value's parent and then modifying a sibling is not
allowed.  In addition, whenever we perform an iterative update
traversing a number of nodes, we require that \emph{the result of an
  iterative update is independent of the order in which the nodes are
  updated}.  We call this the \emph{traversal-order independence
  property}.

To ensure isolation of side effects and traversal-order independence,
it is sufficient to restrict the XPath expressions that can be used to
select foci.  Specifically, only the child axis\footnote{The attribute
  axis can also be handled easily, but the descendant, parent, and
  sibling axes seem nontrivial to handle.} is allowed, and absolute
paths starting with $/$ cannot be used to backtrack to the root of the
document in order to begin iterating over some other part.  This
ensures that only descendants of a given focused value can be selected
as the new focus and that a selection contains no overlapping parts.
Consequently, the side effects of an update are confined to the
subtrees of its focus, and the result of an iteration is independent of
the traversal order.  This keeps the semantics deterministic and helps
make typechecking feasible.

\subsection{Examples}

Suppose we start an XML database with no pre-loaded data; its type is
$db[\emptyseq]$.  We want to create a small database listing books and
authors.  The following \Flux updates accomplish this:
\begin{verbatim}
U1 : INSERT AS LAST INTO db VALUE books[];
     INSERT AS LAST INTO db VALUE authors[]
\end{verbatim}
After this update, the database has type
\begin{verbatim}
books[],authors[]
\end{verbatim}

Suppose we want to load some XML data into the database.  Since XML
text is included in XQuery's expression language, we can just do the
following:
\begin{verbatim}
U2 : INSERT INTO books VALUE 
     <book><author>Charles Dickens</author>
           <title>A Tale of Two Cities</title>
           <year>1858</year></book>
     <book><author>Lewis Carroll</author>
           <title>Alice in Wonderland</title>
           <year>??</year></book>;
     INSERT INTO authors VALUE 
     <author><name>Charles Dickens</name>
             <born>1812</born>
             <died>1870</died></author>
     <author><name>Lewis Carroll</name>
             <born>1832</born>
             <died>1898</died></author>
\end{verbatim}
This results in a database with type
\begin{verbatim}
books[ book[author[string],title[string],
            year[string]]* ],
authors[ author[name[string],born[string],
                died[string]]* ]
\end{verbatim}

The data we initially inserted had some missing dates. We can fill
these in as follows:
\begin{verbatim}
U3 : UPDATE $x AS books/book BY
       REPLACE IN year WITH "1859"
     WHERE $x/title/text() = "A Tale of Two Cities"
U4 : UPDATE $x AS books/book BY
       REPLACE IN year WITH "1865"
     WHERE $x/title/text() = "Alice in Wonderland"
\end{verbatim}
Note that here, we use an XQuery expression \verb|$x/name/text()| for
the $\kWHERE$-clause.  Both updates leave the structure of the database unchanged.

We can add an element to each \verb|book| in \verb|books| as follows:
\begin{verbatim}
U5 : INSERT AS LAST INTO books/book 
     VALUE publisher["Grinch"]
\end{verbatim}
After \verb|U5|, the books database has type
\begin{verbatim}
books[ book[author[string],title[string],
            year[string],publisher[string]]* ]
\end{verbatim}
Now perhaps we want to add a co-author; for example, perhaps Lewis
Carroll collaborated on ``Alice in Wonderland'' with Charles
Dickens.  This is not as easy as adding the publisher field to the end
because we need to select a particular node to insert before or after.
In this case we happen to know that there is only one author, so we
can insert after that; however, this would be incorrect if there were
multiple authors, and we would have to do something else (such as
inserting before the title).
\begin{verbatim}
U6 : UPDATE $x AS books/book BY
       INSERT AFTER author
       VALUE <author>Charles Dickens</author>
     WHERE $x/name/text() = "Alice in Wonderland"
\end{verbatim}
Now the \verb|books| part of the database has
the type:
\begin{verbatim}
books[ book[author[string]*,title[string],
            year[string],publisher[string]]* ]
\end{verbatim}
Now that some books have multiple authors, we might want to change the
flat author lists to nested lists:
\begin{verbatim}
U7 : REPLACE $x AS books/book WITH 
     <book><authors>{$x/author}</authors>
           {$x/title}{$x/year}{$x/publisher}</book>
\end{verbatim}
This visits each book and changes its structure so that the authors
are grouped into an \verb|authors| element.  The resulting
\verb|books| subtree has type:
\begin{verbatim}
books[ book[authors[author[string]* ],title[string],
            year[string],publisher[string]]* ]
\end{verbatim}

Suppose we later decide that the publisher field is unnecessary after
all.  We can get rid of it using the following update:
\begin{verbatim}
U8 : DELETE books/book/publisher
\end{verbatim}
The \verb|books| subtree in the result has type
\begin{verbatim}
books[ book[authors[author[string]* ],
            title[string],year[string]]* ]
\end{verbatim}

Now suppose Lewis Carroll retires and we wish to remove all of
his books from the database.
\begin{verbatim}
U9 : DELETE $x AS books/book
     WHERE $x/authors/author/text() = "Lewis Carroll"
\end{verbatim}
This update does not modify the type of the database.  Finally, we can
delete a top-level document as follows:
\begin{verbatim}
U10 : DELETE authors
\end{verbatim}

\subsection{Non-design goals}

There are several things that other proposals for updating XML do that
we make no attempt to do.  We believe that these design choices are
well-motivated for \Flux's intended application area, database
updates.

\textbf{Node identity:} The XQuery data model provides identifiers for
all nodes.  Many XML update proposals take node identities into
account and can use them as to update parts of the tree ``by
reference''.  In contrast, \Flux's semantics is purely value-based.
Although there are currently no examples involving node identity for
XQuery database updates in the W3C's requirements documents
\cite{xquery-update-requirements-w3c-062006}, node identity is
important in other XML update settings such as the W3C's Document
Object Model (DOM).  We believe it is possible to adapt \Flux to a
data model with node identity as long as the identifiers are not used
as mutable references.

\textbf{Pattern matching:} Many transformation/query languages
(e.g.  \cite{hosoya05toplas,clark99xslt}) and some update languages
(e.g.  \cite{DBLP:conf/ssdbm/LiefkeD99,DBLP:conf/ideas/WangLL03})
allow defining transformations by \emph{pattern matching}, that is,
matching tree patterns against the data.  Pattern matching is very
useful for XML transformations in Web programming (e.g. converting an
XML document into HTML), but we believe it is not as important for
typical XML database updates.  We have not considered general pattern
matching in \Flux, in order to keep the type system and operational
semantics as simple as possible.  

\textbf{Side-effects in queries:} Several motivating examples for
\XQueryBang \cite{DBLP:conf/edbtw/GhelliRS06} and XQueryP
\cite{chamberlin06ximep} depend on the ability to perform side-effects
within queries. Examples include logging accesses to particular data
or profiling or debugging an XQuery program.  \Flux cannot be used for
these applications.  However, it is debatable whether adding
side-effects to XQuery is the best way to support logging, profiling,
or debugging for XQuery.

\section{Core language formalization}\labelSec{formalization}

The high-level update language introduced in the last section is
convenient for users, but its operations are complex, overlapping, and
difficult to typecheck.  Just as for XQuery and many other languages,
it is more convenient to define a core language with orthogonal
operations whose semantics and typing rules are simple and transparent, and
then translate the high-level language to the core language.  \footnote{Such
core languages are also typically easier to optimize, though we do not
consider optimization in this paper.}  We first review the XML data
model, regular expression types, and the \muXQ core query language of
\citet{colazzo06jfp}.

\subsection{XML values and regular expression types}
Following \citet{colazzo06jfp}, we distinguish between \emph{tree
  values} $t \in \Tree$, which include strings $w \in \Sigma^*$ (for
some alphabet $\Sigma$), boolean values $\ktrue,\kfalse \in \Bool$,
and singleton trees $n[v]$ where $n \in \Lab$ is a node label; and
\emph{(forest) values} $v \in \Val = \Tree^*$, which are sequences of
tree values:
\[\begin{array}{lrcl}
  \text{Tree values} & t &::=& n[v] \mid w \mid \ktrue \mid \kfalse
  \smallskip\\
  \text{(Forest) values} & v&::=& \emptyseq \mid t,v 
\end{array}\]
We overload the set membership symbol $\in$ for trees and
forests: that is, $t \in v$ means that $t$ is a member of $v$
considered as a list.  Two forest values can be concatenated by
concatenating them as lists; abusing notation, we identify trees $t$
with singleton forests $t,\emptyseq$ and write $v, v'$ for forest
concatenation.  We define a comprehension operation on forest values
as follows:
\begin{eqnarray*}
  {}[f(x) \mid x \in \emptyseq] &=& \emptyseq\\
  {}[f(x) \mid x \in t,v] &=& f(t), [f(x) \mid x \in v]
\end{eqnarray*}
This operation takes a forest $(t_1,\ldots,t_n)$ and a function $f(x)$
from trees to forests and applies $f$ to each tree $t_i$,
concatenating the resulting forests in order.  Comprehensions satisfy
basic monad laws as well as some additional equations (see
\cite{fernandez01icdt}).  We use $=$ for (mathematical) equality of
tree or forest values.

We consider a regular expression type system with structural
subtyping, similar to those considered in several transformation and
query languages for XML
\cite{hosoya05toplas,colazzo06jfp,fernandez01icdt}.
\[\begin{array}{lrcl}
  \text{Atomic types} & \alpha &::=& \kbool \mid \kstring \mid n[\tau]
  \\
  \text{Sequence types} & \tau,\sigma &::=& \alpha \mid \emptyseq \mid \tau |\tau' \mid \tau,\tau' \mid \tau^* \mid X
\end{array}\]
We call types of the form $\alpha \in \Atoms$ \emph{atomic} types (or
sometimes tree or singular types), and types $\tau,\sigma \in \Types$
of all other forms \emph{sequence types} (or sometimes forest or
plural types).  Sequence types are constructed using regular
expression operations such as the empty sequence $\emptyseq$,
alternative choice $\tau | \tau'$, sequential composition $\tau,\tau'$
and iteration (or Kleene star) $\tau^*$.  Type variables $X \in
\TyVar$ denoting recursively defined types are also allowed; these
must be declared in signatures as discussed below.

A value of singular type must always be a sequence of length one (that
is, a tree, string, or boolean); plural types may have values of any length.  There exist
plural types with only values of length one, but which are not
syntactically singular (for example $\kstring | \kbool$).  As usual,
the $+$ and $?$ quantifiers are definable as follows: $\tau^+ =
\tau,\tau^*$ and $\tau^? = \tau|\emptyseq$.

We define \emph{type definitions} and \emph{signatures} as follows:
\[\begin{array}{lrcl}
\text{Type definitions} &  \tau_0 &::=& \alpha \mid \emptyseq \mid \tau_0 |\tau_0' \mid \tau_0,\tau_0' \mid \tau_0^*\\
  \text{Type signatures} &E & ::= & \cdot \mid E,\typedecl{X}{\tau_0}
\end{array}\]
Type definitions $\tau_0$ are types with no top-level variables (that
is, every variable is enclosed in a $n[-]$ context).  A signature
$E$ is well-formed if all type variables appearing in definitions are
also declared in $E$.  Given a well-formed signature $E$, we write
$E(X)$ for the definition of $X$.  A type $\tau$ denotes the set of
values $\SB{\tau}_E$, defined as follows.
\[\begin{array}{l}
  \begin{array}{rcl}
    \SB{\kstring}_E &=& \Sigma^*\smallskip\\
    \SB{\kbool}_E &=& \Bool\smallskip\\
    \SB{\emptyseq}_E &=& \{\emptyseq\}
\end{array}\quad
\begin{array}{rcl}
    \SB{n[\tau]}_E &=& \{n[v] \mid v \in \SB{\tau}_E\}\smallskip\\
    \SB{\tau|\tau'}_E &=& \SB{\tau}_E \cup \SB{\tau'}_E\smallskip\\
    \SB{X}_E &=& \SB{E(X)}
  \end{array}
  \smallskip\\
  \begin{array}{rcl}
    \SB{\tau,\tau'}_E &=& \{v, v' \mid v \in \SB{\tau}_E,v' \in \SB{\tau'}_E\}
    \smallskip\\
    \SB{\tau^*}_E &=& \bigcup_{n=0}^\infty \{v_1, \ldots, v_n \mid v_1 ,\ldots,v_n \in \SB{\tau}_E\}\\
  \end{array}
\end{array}
\]
Formally, $\SB{\tau}_E$ is defined by a straightforward least fixed
point construction which we omit (see e.g.~\cite{hosoya05toplas}).
Henceforth, we treat $E$ as fixed and define $\SB{\tau} \defeq
\SB{\tau}_{E}$.  This semantics validates standard identities such as
associativity of ',' ($\SB{(\tau_1,\tau_2),\tau_3} =
\SB{\tau_1,(\tau_2,\tau_3)}$), unit laws ($\SB{\tau,\emptyseq} =
\SB{\tau} = \SB{\emptyseq,\tau}$), and idempotence of '*'
($\SB{(\tau^*)^*} = \SB{\tau^*}$).

A type $\tau_1$ is a \emph{subtype} of $\tau_2$ ($\tau_1 \subty
\tau_2$), by definition, if $\SB{\tau_1} \subseteq \SB{\tau_2}$.  The
use of regular expressions (including untagged unions) for XML typing
poses a number of problems for subtyping and typechecking which have
been resolved in previous work on XDuce~\cite{hosoya05toplas}.  Our
types are essentially the same as those used in XDuce, so subtyping
reduces to XDuce subtyping; although this problem is EXPTIME-complete
in general, the algorithm of \citet{hosoya05toplas} is well-behaved in
practice.  Therefore, we shall not give explicit inference rules for
checking or deciding subtyping, but treat it as a ``black box''.

\subsection{Core query language}
Because \Flux uses queries for insertion, replacement, and
conditionals, we need to introduce a query language and define its
semantics before doing the same for \Flux.  In our implementation, we
use a variant of the \muXQ core language introduced by
\citet{colazzo06jfp}, which has the following syntax:
\begin{eqnarray*}
  e &::=& \emptyseq \mid e,e' \mid n[e] \mid w \mid x \mid \letin{x=e}{e'}\\
  &\mid& \ktrue \mid \kfalse\mid \ifthenelse{c}{e}{e'} \mid e \eq e'\\
  &\mid & \bar{x} \mid \bar{x}/\kchild \mid e::n  \mid \forreturn{\bar{x} \in e}{e'}
\end{eqnarray*}
We follow the convention in \cite{colazzo06jfp} of using $\bar{x}$ for
variables introduced by $\kfor$, which are always bound to tree
values; ordinary variables $x$ may be bound to any value.

An \emph{environment} is a pair of functions $\gamma : (\Var \to
\Val)\times(\TVar \to \Tree)$.  Abusing notation, we write $\gamma(x)$
for $\pi_1(\gamma)(x)$ and $\gamma(\bar{x})$ for
$\pi_2(\gamma)(\bar{x})$; similarly, $\gamma[x:=v]$ and
$\gamma[\bar{x}:=t]$ denote the corresponding environment updating
operations.  The semantics of queries is defined via the large-step
operational semantics judgment $\eval{\gamma}{e}{v}$, meaning ``in
environment $\gamma$, expression $e$ evaluates to value $v$''.
\confonly{The contributions of this paper do not require detailed
  understanding of the query language semantics, so the rules are
  relegated to the technical report~\cite{flux-tr}.}  \tronly{The
  contributions of this paper do not require detailed understanding of
  the query language, so the rules are relegated to the appendix.}  We
omit recursive queries but they can be added without difficulty.

\begin{figure*}[tb]
  \fbox{$\evalu{\gamma}{v}{s}{v'}$}
  \[\small\begin{array}{c}
    \infer{\evalu{\gamma}{v}{\kskip}{v}}{}
    \quad
    \infer{\evalu{\gamma}{v}{s;s'}{v_2}}
    {\evalu{\gamma}{v}{s}{v_1} & 
      \evalu{\gamma}{v_1}{s'}{v_2}}
    \quad
    \infer{\evalu{\gamma}{v}{\ifthenelse{e}{s_1}{s_2}}{v'}}
    {\eval{\gamma}{e}{\ktrue}
      & \evalu{\gamma}{v}{s_1}{v'}}
    \quad
    \infer{\evalu{\gamma}{v}{\ifthenelse{e}{s_1}{s_2}}{v'}}
    {\eval{\gamma}{e}{\kfalse}
      & \evalu{\gamma}{v}{s_2}{v'}}
    \smallskip\\
    \infer{\evalu{\gamma}{v_1}{\letin{x=e}{s}}{v_2}}
    {\eval{\gamma}{e}{v}
      &
      \evalu{\gamma[x:=v]}{v_1}{s}{v_2}
    } 
    \quad
    \infer{\evalu{\gamma}{\emptyseq}{\kinsert~e}{v}}
    {\eval{\gamma}{e}{v}}
    \quad\infer{\evalu{\gamma}{v}{\kdelete}{\emptyseq}}
    {}
    \quad 
    \infer{\evalu{\gamma}{n'[v]}{\krename~n}{n[v]}}
    {}
    \smallskip\\ 
    \infer{\evalu{\gamma}{v}{\snapshot{x}{s}}{v'}}
    {\evalu{\gamma[x:=v]}{v}{s}{v'}}
    \quad\infer{\evalu{\gamma}{t}{\phi?s}{v}}
    {t \in \SB{\phi} & \evalu{\gamma}{t}{s}{v}}
    \quad
    \infer{\evalu{\gamma}{t}{\phi?s}{t}}
    {t \not\in \SB{\phi}}
    \quad
    \infer{\evalu{\gamma}{n[v]}{\kchildren[s]}{n[v']}}
    {\evalu{\gamma}{v}{s}{v'}}
    \smallskip\\
    \infer{\evalu{\gamma}{v}{\kleft[s]}{v',v}}
    {\evalu{\gamma}{\emptyseq}{s}{v'}}
    \quad
    \infer{\evalu{\gamma}{v}{\kright[s]}{v,v'}}
    {\evalu{\gamma}{\emptyseq}{s}{v'}}
    \quad
    \infer{\evalu{\gamma}{t_1,v_2}{\kiter[s]}{v_1',v_2'}}
    {\evalu{\gamma}{t_1}{s}{v_1'}
      & 
      \evalu{\gamma}{v_2}{\kiter[s]}{v_2'}}
    \quad\infer{\evalu{\gamma}{\emptyseq}{\kiter[s]}{\emptyseq}}{}
\smallskip\\
\infer{\evalu{\gamma}{v}{P(\vec{e})}{v'}}{
\procdecl{P(\vec{x}:\vec{\tau})}{\tau_1}{\tau_2}\defeq s \in \Delta & 
\eval{\gamma}{e_1}{v_1} & 
\cdots & 
\eval{\gamma}{e_n}{v_n} & 
\evalu{\gamma[x_1:=v_1,\ldots,x_n:=v_n]}{v}{s}{v'}}
  \end{array}\]
  \caption{Operational semantics of
    updates.}\labelFig{lux-core-semantics}
\end{figure*}

\subsection{Core update language}

We now introduce the core \Flux update language, which includes
statements $s \in \Stmt$, tests $\phi \in \Test$, and directions $d
\in \Dir$:
\begin{eqnarray*}
  s &::=& \kskip \mid s;s' \mid \ifthenelse{e}{s}{s'} \mid\letin{x=e}{s} \\
  &\mid & \kinsert~e \mid \kdelete \mid \krename~n \\
  &\mid& \snapshot{x}{s}  \mid \phi?s \mid d[s] \mid P(\vec{e})\\
  \phi &::=& n \mid \knode() \mid \ktext()\\
  d &::=& \kleft\mid \kright \mid \kchildren \mid \kiter
\end{eqnarray*}
Here, $P$ denotes an \emph{update procedure} name.  Procedures are
defined via declarations $P(\vec{x}:\vec{\tau}) : \tau_1 \To \tau_2
\defeq s$, meaning $P$ takes parameters $\vec{x}$ of types $\tau$ and
changes a database of type $\tau_1$ to one of type $\tau_2$.  We
collect these declarations into a set $\Delta$, which we take to be
fixed throughout the rest of the paper.  Procedures may be recursive.

Updates include standard constructs such as the no-op $\kskip$,
sequential composition, conditionals, and $\klet$-binding.  Recall
that updates work by \emph{focusing} on selected parts of the mutable
store.  The basic update operations include insertion $\kinsert~e$,
which inserts a value provided the focus is the empty sequence;
deletion $\kdelete$, which deletes the focus (replacing it with the
empty sequence); and $\krename~n$, which renames the current focused
value (provided it is a singleton tree).  The ``snapshot'' operation
$\snapshot{x}{s}$ binds $x$ to the current focused value and then
applies an update $s$, which may refer to $x$.  There is no way to
refer to the focus of an update within a \muXQ query without using
$\ksnapshot$.  Also, $\ksnapshot$ is \emph{not} equivalent to
\XQueryBang's \verb|snap| operator; $\ksnapshot$ binds $x$ to an
immutable value which can be used in $s$, whereas \verb|snap| forces
execution of pending updates in \XQueryBang.

Updates also include \emph{tests} $\phi?s$ which allow us to examine
the local structure of a tree value and perform an update if the
structure matches.  The node label test $n?s$ checks whether the
focus is of the form $n[v]$, and if so executes $s$, otherwise
is a no-op; the wildcard test $\knode()?s$ only checks that the value is a
singleton tree.  Similarly, $\ktext()?s$ tests whether the focus
is a string.  The $?$ operator binds tightly; for example, $\phi?s;s'
= (\phi?s);s'$.

Finally, updates include \emph{navigation} operators that change the
selected part of the tree and perform an update on the sub-selection.
The $\kleft$ and $\kright$ operators move to the left or right of a
value.  The $\kchildren$ operator shifts focus to the children of a
tree value.  The $\kiter$ operator shifts focus to all of the tree
values in a forest.

We distinguish between \emph{singular} (unary) updates which apply to
tree values and \emph{plural} (multi-ary) updates which apply to
sequences.  Tests $\phi?s$ are always singular.  The $\kchildren$
operator applies a plural update to the children of a single node; the
$\kiter$ operator applies a singular update to all of the elements of
a sequence.  Other updates can be either singular or plural in
different situations.

\refFig{lux-core-semantics} shows the operational semantics of Core
\Flux.  We write $\evalu{\gamma}{v}{s}{v'}$ to indicate that given
environment $\gamma$ and focus $v$, statement $s$ updates $v$
to value $v'$.  The rules for tests are defined in terms of the
following semantic interpretation of tests:
\[\begin{array}{rcl}
 \SB{\ktext()} &=& \Sigma^* 
  \smallskip\\
  \SB{n} &=& \{n[v] \mid v \in \Val\}
  \smallskip\\
  \SB{\knode()} &=&  \Tree
\end{array}\]

Note that we define the semantics entirely in
terms of forest and tree values, without needing to define an explicit
store. This would not be the case if we considered full XQuery, which
includes node identity comparison operations.  However, we believe our
semantics is compatible with allowing node-identity tests in queries.

\begin{theorem}[Update determinism]
  Let $\gamma,v,s, v_1,v_2$ be given such that
  $\evalu{\gamma}{v}{s}{v_1}$ and $\evalu{\gamma}{v}{s}{v_2}$.  Then
  $v_1 = v_2$.
\end{theorem}
\begin{proof}
  Straightforward by induction on the structures of the two
  derivations.  The interesting cases are those for conditionals,
  tests, and iteration, since they are the only statements that have
  more than one applicable rule.  However, in each case, only matching
  pairs of rules are applicable.
\end{proof}

\section{Type system}\labelSec{types}

As noted earlier, certain \emph{singular} updates expect that the
input value is a singleton (for example, $\kchildren$, $n?s$, etc.)
while \emph{plural} updates work for an arbitrary sequence of trees.
Singular updates fail if applied to a sequence.  Our type system
should prevent such run-time failures.  Moreover, as with all XML
transformation languages, we often would like to ensure that when
given an input tree of some type $\tau$, an update is guaranteed to
produce an output tree of some other type $\tau'$.  For example,
updates made by non-privileged users are usually required to preserve
the database schema.

We define a matching relation between tree types and tests: we
say that $\alpha \subty \phi$ if $\SB{\alpha} \subseteq \SB{\phi}$.
This is decidable using the following rules:
\[
\infer{\kstring \subty \ktext()}{}\quad \infer{n[\tau] \subty
  n}{}\quad \infer{\alpha \subty \knode()}{}\]

\begin{figure}[tb]
  \fbox{$\wfupd{\Gamma}{a}{\tau}{s}{\tau'}$}
  \[\small\begin{array}{c}
    \infer{\wfupd{\Gamma}{a}{\tau}{\kskip}{\tau}}{}
    \quad
    \infer{\wfupd{\Gamma}{a}{\tau}{s;s'}{ \tau''}}
    {\wfupd{\Gamma}{a}{\tau}{s}{\tau'} & \wfupd{\Gamma}{a}{\tau'}{s'}{ \tau''}}
    \smallskip\\
    \infer{\wfupd{\Gamma}{a}{\tau}{\ifthenelse{e}{s}{s'}}{\tau_1 | \tau_2}}
    {\wf{\Gamma}{e}{\kbool} & \wfupd{\Gamma}{a}{\tau}{s}{\tau_1} & \wfupd{\Gamma}{a}{\tau}{s'}{\tau_2}}
    \smallskip\\
    \infer{\wfupd{\Gamma}{*}{\emptyseq}{\kinsert~e}{\tau}}
    {\wf{\Gamma}{e}{\tau}}
    \quad
    \infer{\wfupd{\Gamma}{a}{\tau}{\kdelete}{\emptyseq}}
    {}
    \smallskip\\
    \infer{\wfupd{\Gamma}{1}{n'[\tau]}{\krename~n}{n[\tau]}}
    {}
    \quad
    \infer{\wfupd{\Gamma}{a}{\tau_1}{\letin{x=e}{s}}{\tau_2}}
    {\wf{\Gamma}{e}{\tau} & \wfupd{\Gamma,x{:}\tau}{a}{\tau_1}{s}{\tau_2}}
    \smallskip\\
    \infer{\wfupd{\Gamma}{a}{\tau}{\snapshot{x}{s}}{\tau'}}{\wfupd{\Gamma,x{:}\tau}{a}{\tau}{s}{\tau'}}
    \quad
    \infer{\wfupd{\Gamma}{1}{\alpha}{\phi?s}{\tau}}
    {\alpha \subty \phi & \wfupd{\Gamma}{1}{\alpha}{s}{\tau}}
    \smallskip\\
    \infer{\wfupd{\Gamma}{1}{\alpha}{\phi?s}{\alpha}}
    {\alpha \not\subty \phi}
    \quad
    \infer{\wfupd{\Gamma}{1}{n[\tau]}{\kchildren[s]}{n[\tau']}}
    {\wfupd{\Gamma}{*}{\tau}{s}{\tau'}}
    \smallskip\\
    \infer{\wfupd{\Gamma}{a}{\tau}{\kleft[s]}{\tau',\tau}}
    {\wfupd{\Gamma}{*}{\emptyseq}{s}{\tau'}}
    \quad
    \infer{\wfupd{\Gamma}{a}{\tau}{\kright[s]}{\tau,\tau'}}
    {\wfupd{\Gamma}{*}{\emptyseq}{s}{\tau'}}
    \smallskip\\
    \infer{\wfupd{\Gamma}{*}{\tau}{\kiter[s]}{\tau'}}
    {\wfiter{\Gamma}{\tau}{s}{\tau'}}
    \quad
    \infer{\wfupd{\Gamma}{a}{\tau_1}{s}{\tau_2}}
    { 
      \wfupd{\Gamma}{a}{\tau_1}{s}{\tau_2'} & 
      \tau_2' \subty \tau_2}
\smallskip\\
\infer{\wfupd{\Gamma}{a}{\sigma_1'}{P(\vec{e})}{\sigma_2}}{
\begin{array}{l}
  \procdecl{P(\vec{x}:\vec{\tau})}{\sigma_1}{\sigma_2}\defeq s \in \Delta \quad \sigma_1' \subty \sigma_1\\
\wf{\Gamma}{e_1}{\tau_1'} \quad \tau_1' \subty \tau_1 \quad \cdots \quad
\wf{\Gamma}{e_n}{\tau_n'} \quad \tau_n' \subty \tau_n
\end{array}
}
  \end{array}\]
  \fbox{$\wfiter{\Gamma}{\tau}{s}{\tau'}$}
  \[\small\begin{array}{c}
    \infer{\wfiter{\Gamma}{\emptyseq}{s}{\emptyseq}}{}
    \quad
    \infer{\wfiter{\Gamma}{\alpha}{s}{\tau}}
    {\wfupd{\Gamma}{1}{\alpha}{s}{\tau}}
    \quad
    \infer{\wfiter{\Gamma}{\tau_1^*}{s}{\tau_2^*}}
    {\wfiter{\Gamma}{\tau_1}{s}{\tau_2^*}}
    \smallskip\\
    \infer{\wfiter{\Gamma}{\tau_1,\tau_2}{s}{\tau_1',\tau_2'}}
    {\wfiter{\Gamma}{\tau_1}{s}{\tau_1'}
      &
      \wfiter{\Gamma}{\tau_2}{s}{\tau_2'}}
    \smallskip\\
    \infer{\wfiter{\Gamma}{\tau_1|\tau_2}{s}{\tau_1'|\tau_2'}}
    {\wfiter{\Gamma}{\tau_1}{s}{\tau_1'}
      &
      \wfiter{\Gamma}{\tau_2}{s}{\tau_2'}}
\quad
    \infer{\wfiter{\Gamma}{X}{s}{\tau}}
    {\wfiter{\Gamma}{E(X)}{s}{\tau}}
  \end{array}\]
\fbox{$\wfdecl{\Delta}$}
\[\small\begin{array}{c}
\infer{\wfdecl{\emptyset}}{} \quad 
\infer{\wfdecl{\Delta,\procdecl{P(\vec{x}:\vec{\tau})}{\tau_1}{\tau_2}\defeq s}}
{\wfdecl{\Delta} & \wfupd{\vec{x}:\vec{\tau}}{*}{\tau_1}{s}{\tau_2}}
\end{array}\]
\caption{Update, iteration, and declaration
  well-formedness.}\labelFig{update-wf}
\end{figure}

We employ a type system for queries similar to that developed by
\citet{colazzo06jfp}.  We consider type environments $\Gamma$
consisting of sets of bindings $x{:}\tau$ of variables to types and
$\bar{x}{:}\alpha$ of tree variables to atomic types.  (We never need
to bind a tree variable to a sequence type).  As usual, we assume that
variables in type environments are distinct; this convention
implicitly constrains all inference rules.  We write $\SB{\Gamma}$ for
the set of all environments $\gamma$ such that $\gamma(x) \in
\SB{\Gamma(x)}$ and $\gamma(\bar{x}) \in \SB{\Gamma(\bar{x})}$ for all
$x \in dom(\Gamma)$ and $\bar{x} \in dom(\Gamma)$ respectively.

The typing judgment for queries is $\wf{\Gamma}{e}{\tau}$, meaning
\emph{in type environment $\Gamma$, expression $e$ has type $\tau$}.
The typing rules are essentially the same as those
in~\cite{colazzo06jfp}.

The main typing judgment for updates is
$\wfupd{\Gamma}{a}{\tau}{s}{\tau'}$, meaning \emph{in type environment
  $\Gamma$, an $a$-ary update $s$ maps values of type $\tau$ to type
  $\tau'$}.  Here, $a \in \{1,*\}$ is the arity of the update, and
singular update judgments always have $\tau = \alpha$ atomic.  In
addition, we define auxiliary judgments
$\wfiter{\Gamma}{\tau}{s}{\tau'}$ for typechecking iterations and
$\wfdecl{\Delta}$ for typechecking declarations $\Delta$.  The rules
for update well-formedness are shown in \refFig{update-wf}.

\subsection{Discussion}

In many functional languages, and several XML update proposals,
side-effecting operations are treated as expressions that return
$\emptyseq$.  Thus, we could typecheck such updates as expressions of
type $\emptyseq$. This is straightforward provided the types of values
reachable from the free variables in $\Gamma$ do not change; for
example, this is the case for ML-style references.  However, if the
side-effects do change the types of the values of variables, then
$\Gamma$ needs to be updated to take these changes into account.  One
possibility is to typecheck updates using a residuating judgment
$\Gamma \vdash s : \emptyseq \mid \Gamma'$; here, $\Gamma'$ is the
updated type environment reflecting the types of the variables after
update $s$.  This approach quickly becomes complicated, especially if
it is possible for variables to ``alias'', or refer to overlapping
parts of the data.

In \Flux, we take a completely different approach to typechecking
updates.  The judgment $\wfupd{\Gamma}{a}{\tau}{s}{\tau'}$ assigns an
update much richer type information that describes the type of the
updatable context before and after running $s$.  The variables in
$\Gamma$ are immutable, so their types never need to be updated.

The most unusual rules are those involving the $\kiter$, test, and
$\kchildren$, $\kleft/\kright$, and $\kinsert/\krename/\kdelete$
operators.  The following example illustrates how the rules work
for these constructs.  Consider the update:
\[\kiter~[a?\kchildren~[\kiter~[b?~\kright~[\kinsert~c[]]]]]\]
Intuitively, this update inserts a $c$ after every $b$ under a
top-level $a$.  Now consider the input type $a[b[]^*,c[],b[]^*],d[]$.
Clearly, the output type \emph{should} be
$a[(b[],c[])^*,c[],(b[],c[])^*],d[]$.  To see why this is the case,
first note that the following can be derived for any $\tau,\tau',s$:
\[\infer{\wfupd{}{*}{a[\tau],d[]}{\kiter~[a?s]}{a[\tau'],d[]}}
{\wfupd{}{1}{a[\tau]}{s}{a[\tau']}}\]
Using the rule for $\kchildren$, we can see that it suffices to check
that $\kiter~[b?\kright~[\kinsert~c[]]]$ maps type $b[]^*,c[],b[]^*$ to
$(b[],c[])^*,c[],(b[],c[])^*$.  This is also an instance of a
derivable rule
\[\infer{\wfupd{}{*}{b[]^*,c[],b[]^*}{\kiter~[b?s]}{\tau^*,c[],\tau^*}}
{\wfupd{}{1}{b[]}{s}{\tau}}\]
Hence, we now need to show only that $\kright~[\kinsert~c[]]$ maps type
$b[]$ to $b[],c[]$, which is immediate:
\[
\infer{\wfupd{}{1}{b[]}{\kright~[\kinsert~c[]]}{b[],c[]}}
{\infer{\wfupd{}{*}{\emptyseq}{\kinsert~c[]}{c[]}}
  {\infer{\wf{}{c[\emptyseq]}{c[\emptyseq]}}{\hyp{\wf{}{\emptyseq}{\emptyseq}}}}}
\]

\subsection{Metatheory}
We take for granted the following type soundness property for queries
(this was proved for \muXQ in \citet{colazzo06jfp}).
\begin{theorem}[Query soundness]\labelThm{query-soundness}
  If $\wf{\Gamma}{e}{\tau}$ and $\gamma \in \SB{\Gamma}$ then
  $\eval{\gamma}{e}{v}$ implies $v \in \SB{ \tau}$.
\end{theorem}
\confonly{The corresponding result also holds for updates, by a straightforward
structural induction argument (presented in the technical
report~\cite{flux-tr}):}
\tronly{The corresponding result also holds for updates, by a straightforward
structural induction argument (presented in the appendix):}
\begin{theorem}[Update soundness]\labelThm{update-soundness}
  Assume $\wfdecl{\Delta}$ holds.
  \begin{enumerate}
  \item If $\wfupd{\Gamma}{a}{\tau}{s}{\tau'}$, $v \in \SB{\tau}$, and
    $\gamma \in \SB{\Gamma}$, then $\evalu{\gamma}{v}{s}{v'}$ implies $v'
    \in \SB{ \tau'}$.
  \item If $\wfiter{\Gamma}{\tau}{s}{\tau'}$, $v \in \SB{\tau}$, and
    $\gamma \in \SB{\Gamma}$, then $\evalu{\gamma}{v}{\kiter[s]}{v'}$
    implies $v' \in \SB{ \tau'}$.
  \end{enumerate}
\end{theorem}
Moreover, typechecking is decidable for both \muXQ and \Flux in the
presence of the subsumption rules~\cite{cheney08esop}.

\section{Normalization}\labelSec{normalization}

\begin{figure*}[tb]
\[\small
\begin{array}{c}
\begin{array}{rcl}
\nstmt{u} &=& \nupd{u}\\
\nstmt{\IFTHEN{e}{s}} &=& \ifthenelse{e}{\nstmt{s}}{\kskip}\\
\nstmt{s_1;s_2} &=& \nstmt{s_1};\nstmt{s_2}\\
\nstmt{\LETIN{x=e}{s}} &=& \letin{x=e}{\nstmt{s}}
\bigskip\\
\npath{.}(s) &=& s\\
\npath{p/p'}(s) &=& \npath{p}(\npath{p'}(s))\\
\npath{\phi}(s) &=& \kchildren[\kiter[\phi?s]]\\
\npath{p[e]}(s) &=& \npath{p}(\ifthenelse{e}{s}{\kskip})\\
\npath{x~\kAS~p}(s) &=& \npath{p}(\snapshot{x}{s})
\end{array}
\end{array}
\!\!\!\!\!
\!\!\!\!\!
\begin{array}{rcl}
\nupd{\insertvalue{\kBEFORE}{p}{e}} &=& \npath{p}(\kleft[\kinsert~e])\\
\nupd{\insertvalue{\kAFTER}{p}{e}} &=& \npath{p}(\kright[\kinsert~e])\\
\nupd{\insertinto{\kFIRST}{p}{e}} &=& \npath{p}(\kchildren[\kleft[\kinsert~e]])\\
\nupd{\insertinto{\kLAST}{p}{e}} &=& \npath{p}(\kchildren[\kright[\kinsert~e]])\\
\nupd{\delete{p}} &=& \npath{p}(\kdelete)\\
\nupd{\deletefrom{p}} &=& \npath{p}(\kchildren[\kdelete])\\
\nupd{\rename{p}{n}} &=& \npath{p}(\krename~n)\\
\nupd{\replace{p}{e}} &=& \npath{p}(\kdelete;\kinsert~e)\\
\nupd{\replacein{p}{e}} &=& \npath{p}(\kchildren[\kdelete;\kinsert~ e])\\
\nupd{\update{p}{s}} &=& \npath{p}(\nstmt{s})
\end{array}\]
\caption{Source update normalization}\labelFig{norm-upd}
\end{figure*}

There is a significant gap between the high-level \Flux language we
presented in \refSec{examples} and the core language in the previous
section.  In this section, we formalize a translation from the source
language presented in \refSec{examples} to Core \Flux.  In XQuery,
this kind of translation is called \emph{normalization}.  We define
three normalization functions called \emph{path expression
  normalization} $\elab[\Path]{-}(-)$, \emph{update statement
  normalization} $\elab[\Stmt]{-}$, and \emph{simple update
  normalization} $\elab[\Upd]{-}$.  These functions are defined in
\refFig{norm-upd}.

Path expression normalization takes an extra parameter, which must be
a core \Flux update; that is, $\elab[\Path]{p}(s)$ normalizes a path
$p$ by expanding it to an expression which navigates to $p$ and then
does $s$.  Compound statement normalization is straightforward.  Simple
updates are first normalized by translation to $p$.  We omit the cases
needed to handle $\kWHERE$-clauses; however, they can be handled by
the existing translation if we consider e.g. $\replace{p}{e}~\kWHERE~c$
to be an abbreviation for $\replace{p[c]}{e}$, etc.  In particular,
note that the translation places both $c$ and $e$ into the scope of all
variables declared in $p$.

Since the translation rules cover all cases and are orthogonal, it is
straightforward to see that the normalization functions are total
functions from the source language to Core \Flux.

\subsection{Typechecking source updates}

\begin{figure}
  \fbox{$\wfupdstmt{\Gamma}{\tau}{s}{\tau'}$}
  \[\begin{array}{c}
    \infer{\wfupdstmt{\Gamma}{\tau}{s_1;s_2}{\tau''}}{\wfupdstmt{\Gamma}{\tau}{s_1}{\tau'} & \wfupdstmt{\Gamma}{\tau'}{s_2}{\tau''}}
    \smallskip\\
    \infer{\wfupdstmt{\Gamma}{\tau}{\IFTHEN{e}{s}}{\tau|\tau'}}
    {\wf{\Gamma}{e}{\kbool} & \wfupdstmt{\Gamma}{\alpha}{s}{\tau'}}
    \smallskip\\
    \infer{\wfupdstmt{\Gamma}{\tau}{\LETIN{x=e}{s}}{\tau'}}{\wf{\Gamma}{e}{\tau_0} & \wfupdstmt{\Gamma,x:\tau+_0}{\tau}{s}{\tau'}}  
\smallskip\\
\infer{\wfupdstmt{\Gamma}{\alpha}{u}{\alpha'}}{\wfupdsimp{\Gamma}{\alpha}{u}{\alpha'}}
  \end{array}\]  
\caption{Typechecking rules for compound updates}\labelFig{selected-source-tc-stmt}
\end{figure}

Normalization complicates type-error reporting, since we cannot always
easily explain why the translation of an update fails to typecheck in
source-level terms familiar to the user.  We therefore also develop a
type system for the source language that is \emph{both sound and
  complete} with respect to core \Flux typechecking.  This type system
can therefore be used to report type errors to users in terms of the
source language.  
We assume that query subexpressions $e$ have already been normalized
to \muXQ according to the standard XQuery normalization rules
\cite{xquery-semantics-w3c-20070123}.  The problem of typechecking
unnormalized XQuery expressions is an orthogonal issue (and one that
has to our knowledge not been addressed).

Typechecking source-level updates is challenging because simple
updates may change the types of many parts of the document
simultaneously, depending on the structure of the path $p$.  In
contrast, core \Flux updates are easy to typecheck because they break
the corresponding navigation, selection, and modification of types
into small, manageable steps.

To deal with the non-local nature of source updates, we employ
\emph{type variables} $Z$ and \emph{context-tagged type substitutions}
$\Theta$.  These substitutions are defined as follows:
\[\Theta ::= \emptyset \mid \Theta,Z \mapsto (\Gamma\tri \tau)\]
We distinguish the type variables $Z$ we will use here for
typechecking source updates from the type variables $X$ used in
recursive type definitions $E$; we refer to the latter as
\emph{defined type variables}.  We require the bindings $Z$ in
$\Theta$ to be unique.

We often treat substitutions $\Theta$ as sets or finite maps and in
particular write $\Theta \uplus \Theta'$ for the context-tagged
substitution resulting from taking the union of the bindings in
$\Theta$ and $\Theta'$, provided their domains are disjoint.  We also
write $\tau\subst{\Theta}$ for the result of replacing each occurrence
of an undefined type variable in $\tau$ with its binding in $\Theta$.
Moreover, we write $\Theta\subst{\Theta'}$ for the result of applying
$\Theta'$ to each type in $\Theta$, again ignoring contexts.
Substitution application ignores the contexts $\Gamma$; they are only
used to typecheck updates within the scope of a path.  We consider the
free type variables of $\Theta$ to be the free variables of $\Gamma$
and $\tau$ in the bindings $\Gamma \tri\tau$.

To typecheck a simple update such as $\delete{a/b}$ against an atomic
type such as $a[b[],c[]]$, we proceed as follows:
\begin{enumerate}
\item First \emph{match $p$ against the input type, and
    split the type $\alpha$ of the document into a pair
    $(\alpha',\Theta)$ such that $\alpha = \alpha'\subst{\Theta}$.}

  For example, $a[b[],c[]] = a[Z,c[]]\subst{Z \mapsto \cdot \tri b[]}$.
\item Next \emph{modify $\Theta$ according to the update operation to
    obtain $\Theta'$.}  For this we use the Core \Flux type system to
  update each binding in $\Theta$.  This is only a convenience.
  
  Continuing the example, for a deletion we update $Z \mapsto \cdot \tri b[]$ to
  $Z \mapsto \cdot \tri \emptyseq$.
\item Finally, \emph{apply $\Theta'$ to $\alpha'$ to get the desired
    final type after the update}.

For example, applying $a[Z,c[]]\subst{Z \mapsto \cdot \tri \emptyseq}$ we
get $a[(),c[]]$, as desired (this is equivalent to $a[c[]]$).
\end{enumerate}

Figures \ref{fig:selected-source-upd-tc} and
\ref{fig:selected-source-path-tc} show selected typechecking judgments
for simple updates and paths.  
We introduce auxiliary judgments such as path filter
checking ($\wfupdfilt{\Gamma}{\tau}{\phi}{\tau'}{\Theta}$, shown in
\refFig{selected-source-path-filt-tc}), simultaneous core statement
checking ($\wfupds{\Theta}{s}{\Theta'}$), and simultaneous path
checking ($\wfupdpaths{\Theta}{p}{\Theta'}{\Theta''}$.  The 

For many of the typechecking judgments, we also need to typecheck an
expression against all of the bindings of a context-tagged
substitution $\Theta$. We therefore introduce several
\emph{simultaneous typechecking} judgments.  \confonly{The full system, including the
(straightforward) compound statement typechecking judgment
$\wfupdstmt{\Gamma}{\alpha}{s}{\alpha'}$ and all auxiliary judgments,
is shown in the technical report~\cite{flux-tr}.}
\tronly{The full system, including the
(straightforward) compound statement typechecking judgment
$\wfupdstmt{\Gamma}{\alpha}{s}{\alpha'}$ and all auxiliary judgments,
is shown in the appendix.}

The simple update typechecking rules each follow the procedure
outlined above.  The path typechecking rules match $p$ against the
input type $\alpha$ as described in step 2 above.  Note that paths may
bind variables, and the same variable may be bound to different types
in different cases; this is why we need to include contexts $\Gamma$
in each binding of the substitutions $\Theta$.

The following operation $\Theta \oplus x$ is used to typecheck
$x~\kAS~p$; it adds the binding $x:\tau$ to each binding $\Gamma \tri
\tau$ in $\Theta$.
\[\small
\begin{array}{rcl}
  \emptyset \oplus x &=& \emptyset \smallskip\\
  (\Theta,Z\mapsto (\Gamma \tri \tau)) \oplus x &=& \Theta \oplus x, Z \mapsto (\Gamma,x:\tau \tri \tau)
\smallskip\\
\end{array}\]
The typechecking rule for conditional paths $p[e]$ is slightly subtle.
After typechecking $p$, we obtain a pair $(\alpha',\Theta)$ that
splits $\alpha$ into an unchanged part $\alpha'$ and a substitution
$\Theta$ showing where changes may occur.  Since we do not know
whether $e$ will hold, we must adjust $\alpha'$ by replacing each
occurrence of a variable $Z$ with $Z|\Theta(Z)$.  This is accomplished
using the substitution $\maybe{\Theta}$:
\[\small
\begin{array}{rcl}
  \maybe{\emptyset} &=& \emptyset
\smallskip\\
  \maybe {\Theta,Z\mapsto (\Gamma \tri \alpha)} &=& \maybe{\Theta}, Z \mapsto (\Gamma \tri Z|\alpha)
\end{array}\]

\begin{figure}
\fbox{$\wfupdsimp{\Gamma}{\alpha}{u}{\alpha'}$}
  \[\small\begin{array}{c}
    \infer{\wfupdsimp{\Gamma}{\alpha}{\insertvalue{\kBEFORE}{p}{e}}{\alpha'\subst{\Theta'}}}
{\wfupdpath{\Gamma}{\alpha}{p}{\alpha'}{\Theta} & \wfupds{\Theta}{\kleft[\kinsert~e]}{\Theta'}}
\smallskip\\
    \infer{\wfupdsimp{\Gamma}{\alpha}{\insertinto{\kFIRST}{p}{e}}{\alpha'\subst{\Theta'}}}
{\wfupdpath{\Gamma}{\alpha}{p}{\alpha'}{\Theta} & \wfupds{\Theta}{\kchildren[\kleft[\kinsert~e]]}{\Theta'}}
\smallskip\\
\infer{\wfupdsimp{\Gamma}{\alpha}{\delete{p}}{\alpha'\subst{\Theta'}}}
{\wfupdpath{\Gamma}{\alpha}{p}{\alpha'}{\Theta} & 
  \wfupds{\Theta}{\kdelete}{\Theta'}}
  \end{array}\]
\caption{Selected typechecking rules for simple updates}\labelFig{selected-source-upd-tc}
  \fbox{$\wfupdpath{\Gamma}{\alpha}{p}{\alpha'}{\Theta}$}
  \[\small\begin{array}{c}
\infer{\wfupdpath{\Gamma}{\alpha}{p/p'}{\alpha'\subst{\Theta_2}}{\Theta_2'}}
{\wfupdpath{\Gamma}{\alpha}{p}{\alpha'}{\Theta_1}
&
\wfupdpaths{\Theta_1}{p'}{\Theta_2}{\Theta_2'}}
\smallskip\\
    \infer{\wfupdpath{\Gamma}{\alpha}{.}{\alpha}{\emptyset}}{}
\quad
\infer{\wfupdpath{\Gamma}{\alpha}{p[e]}{\alpha'\subst{\maybe{\Theta}}}{\Theta}}
{\wfupdpath{\Gamma}{\alpha}{p}{\alpha'}{\Theta} & 
\wfs{\Theta}{e}{\kbool}}
\smallskip\\
\infer{\wfupdpath{\Gamma}{\alpha}{x ~\kAS~p}{\alpha'}{\Theta\oplus x}}
{\wfupdpath{\Gamma}{\alpha}{p}{\alpha'}{\Theta}}
\quad
\infer{\wfupdpath{\Gamma}{n[\tau]}{\phi}{n[\tau']}{\Theta}}
{\wfupdfilt{\Gamma}{\tau}{\phi}{\tau'}{\Theta}}
  \end{array}\]
  \caption{Typechecking rules for
    paths}\labelFig{selected-source-path-tc}
  \fbox{$\wfupdfilt{\Gamma}{\tau}{\phi}{\tau'}{\Theta}$}
\[\begin{array}{c}
  \infer{\wfupdfilt{\Gamma}{\emptyseq}{\phi}{\emptyseq}{\emptyset}}{}
  \quad
  \infer{\wfupdfilt{\Gamma}{\alpha}{\phi}{\alpha}{\emptyset}}{\alpha \not\subty \phi}
  \smallskip\\
  \infer{\wfupdfilt{\Gamma}{\alpha}{\phi}{Z}{Z \mapsto (\Gamma \tri\alpha)}}{\alpha \subty \phi & \text{$Z$ fresh}}
  \smallskip\\
  \infer{\wfupdfilt{\Gamma}{\tau_1,\tau_2}{\phi}{(\tau_1',\tau_2')}{\Theta_1 \uplus \Theta_2}}
{\wfupdfilt{\Gamma}{\tau_1}{\phi}{\tau_1'}{\Theta_1}
& 
\wfupdfilt{\Gamma}{\tau_2}{\phi}{\tau_2'}{\Theta_2}}
  \smallskip\\
  \infer{\wfupdfilt{\Gamma}{\tau_1|\tau_2}{\phi}{\tau_1'|\tau_2'}{\Theta_1 \uplus \Theta_2}}
{\wfupdfilt{\Gamma}{\tau_1}{\phi}{\tau_1'}{\Theta_1}
& 
\wfupdfilt{\Gamma}{\tau_2}{\phi}{\tau_2'}{\Theta_2}}
  \smallskip\\
  \infer{\wfupdfilt{\Gamma}{\tau_1^*}{\phi}{\tau_2^*}{\Theta}}
{\wfupdfilt{\Gamma}{\tau_1}{\phi}{\tau_2}{\Theta}}
\quad
\infer{\wfupdfilt{\Gamma}{X}{\phi}{\tau'}{\Theta}}
{\wfupdfilt{\Gamma}{E(X)}{\phi}{\tau'}{\Theta}}
\end{array}\]
  \caption{Typechecking rules for path filters}\labelFig{selected-source-path-filt-tc}
  \fbox{$\wfs{\Theta}{e}{\tau}$}
  \[\begin{array}{c}
    \infer{\wfs{\emptyset}{e}{\tau}}{}
\quad
    \infer{\wfs{\Theta,Z \mapsto (\Gamma \tri \alpha)}{e}{\tau}}
    {\wfs{\Theta}{e}{\tau} & 
      \wf{\Gamma}{e}{\tau}}
  \end{array}\]
  \fbox{$\wfupds{\Theta}{s}{\Theta'}$}
  \[\begin{array}{c}
    \infer{\wfupds{\emptyset}{s}{\emptyset}}{}
\quad
    \infer{\wfupds{\Theta,Z \mapsto (\Gamma \tri \alpha)}{s}{\Theta',Z \mapsto (\Gamma \tri \tau)}}{\wfupds{\Theta}{s}{\Theta'} &\wfupd{\Gamma}{1}{\alpha}{s}{\tau}}
  \end{array}\]
  \fbox{$\wfupdstmts{\Theta}{s}{\Theta'}$}
  \[\small\begin{array}{c}
    \infer{\wfupdstmts{\emptyset}{s}{\emptyset}}{}
\quad
    \infer{\wfupdstmts{\Theta,Z \mapsto (\Gamma \tri \tau)}{s}{\Theta',Z \mapsto (\Gamma \tri \tau')}}{\wfupdstmts{\Theta}{s}{\Theta'} & \wfupdstmt{\Gamma}{\tau}{s}{\tau'}}
  \end{array}\] 
  \fbox{$\wfupdpaths{\Theta}{p}{\Theta'}{\Theta''}$}
  \[\begin{array}{c}
    \infer{\wfupdpaths{\emptyset}{p}{\emptyset}{\emptyset}}{}
\smallskip\\
    \infer{\wfupdpaths{\Theta,Z \mapsto (\Gamma \tri \alpha)}{p}{\Theta',Z \mapsto (\Gamma \tri \alpha')}{\Theta'' \uplus \Theta'''}}
    {\wfupdpaths{\Theta}{p}{\Theta'}{\Theta''} &
      \wfupdpath{\Gamma}{\alpha}{p}{\alpha'}{\Theta'''}}
  \end{array}\] 
 \caption{Simultaneous  typechecking judgments}\labelFig{selected-simult-source-tc}
\end{figure}


\subsection{Metatheory}

Whenever we translate between two typed languages, we would like to
know whether the translation is \emph{sound} (i.e.,
\emph{type-preserving}).  This ensures that if we typecheck the
expression in the source language then its translation will also
typecheck.  \footnote{In an implementation, one often wants to
  re-typecheck after translation anyway as a sanity check for the
  translator.}  Conversely, if the source language expression fails to
typecheck, it is preferable to report the error in terms of the source
language using the source type system.  We have established that the
translation is indeed sound:

\begin{theorem}[Soundness]
  Assume $\Gamma,\tau,\tau'$ have no free type variables $Z$.  Then if
  $\wfupdstmt{\Gamma}{\tau}{s}{\tau'}$ then
  $\wfupd{\Gamma}{*}{\tau}{\nstmt{s}}{\tau'}$.
\end{theorem}

Conversely, another concern is that the source-level type system might
be too restrictive.  Are there source-level expressions whose
\emph{translations} are well-formed, but which are not well-formed in
the source-level system?  This is the question of \emph{completeness},
that is, whether the translation \emph{reflects} typability.  If this
completeness property did not hold, this would indicate that the
source type system could be made more expressive.  Fortunately,
however, completeness does hold:

\begin{theorem}[Completeness]
  Assume $\Gamma,\tau,\tau'$ have no free type variables $Z$.  Then if
  $\wfupd{\Gamma}{*}{\tau}{\nstmt{s}}{\tau'}$ then
  $\wfupdstmt{\Gamma}{\tau}{s}{\tau'}$.
\end{theorem}

\section{Path-errors  and dead-code analysis}\labelSec{path-analysis}

\begin{figure*}[tb]
  \fbox{$\wfupdeff{\Gamma}{a}{\tau}{s}{\tau'}{L}$}
  \[\small\begin{array}{c}
     \infer{\wfupdeff{\Gamma}{a}{\tau}{\kskip_l}{\tau}{\{l\}}}{}
    \quad
    \infer{\wfupdeff{\Gamma}{a}{\tau}{((s_1)_{l_1};(s_2)_{l_2})_l}{ \tau''}{(L_1 \cup L_2)[l_1,l_2 {\To} l]}}
    {\wfupdeff{\Gamma}{a}{\tau}{(s_1)_{l_1}}{\tau'}{L_1} & \wfupdeff{\Gamma}{a}{\tau'}{(s_2)_{l_2}}{ \tau''}{L_2}}
   \quad
    \infer{\wfupdeff{\Gamma}{a}{\tau}{(\snapshot{x}{s_{l'}})_l}{\tau'}{L[l' {\To} l]}}
    {\wfupdeff{\Gamma,x{:}\tau}{a}{\tau}{s_{l'}}{\tau'}{L}}
    \smallskip\\
    \infer{\wfupdeff{\Gamma}{a}{\tau}{(\ifthenelse{e}{(s_1)_{l_1}}{(s_2)_{l_2}})_l}{\tau_1 | \tau_2}{(L_1 \cup L_2)[l_1,l_2 {\To} l]}}
    {\wf{\Gamma}{e}{\kbool} & 
      \wfupdeff{\Gamma}{a}{\tau}{(s_1)_{l_1}}{\tau_1}{L_1} & 
      \wfupdeff{\Gamma}{a}{\tau}{(s_2)_{l_2}}{\tau_2}{L_2}}
    \quad    
    \infer{\wfupdeff{\Gamma}{a}{\tau_1}{(\letin{x=e}{s_{l'}})_l}{\tau_2}{L[l' {\To} l]}}
    {\wf{\Gamma}{e}{\tau} & \wfupdeff{\Gamma,x{:}\tau}{a}{\tau_1}{s}{\tau_2}{L}}
    \smallskip\\
    \infer{\wfupdeff{\Gamma}{*}{\emptyseq}{(\kinsert~e)_l}{\tau}{\{l \mid \tau \subty \emptyseq\}}}
    {\wf{\Gamma}{e}{\tau}}
    \quad
    \infer{\wfupdeff{\Gamma}{a}{\tau}{\kdelete_l}{\emptyseq}{\{l \mid \tau \subty \emptyseq\}}}
    {}
    \quad
    \infer{\wfupdeff{\Gamma}{1}{n'[\tau]}{(\krename~n)_l}{n[\tau]}{\{l \mid n = n'\}}}
    {}
    \smallskip\\
    \infer{\wfupdeff{\Gamma}{1}{\alpha}{(\phi?s_{l'})_l}{\tau}{L[l' {\To} l]}}
    {\alpha \subty \phi & \wfupdeff{\Gamma}{1}{\alpha}{s_{l'}}{\tau}{L}}
    \quad
    \infer{\wfupdeff{\Gamma}{1}{\alpha}{(\phi?s_{l'})_{l}}{\alpha}{\{l\}}}
    {\alpha \not\subty \phi}
    \quad
    \infer{\wfupdeff{\Gamma}{*}{\tau}{\kiter[s_{l'}]_l}{\tau'}{L[l'{\To}l]}}
    {\wfitereff{\Gamma}{\tau}{s_{l'}}{\tau'}{L}}
\quad
\infer{\wfupdeff{\Gamma}{a}{\tau}{P(\vec{e})}{\tau'}{\emptyset}}
{\wfupd{\Gamma}{a}{\tau}{P(\vec{e})}{\tau'}}
    \smallskip\\
    \infer{\wfupdeff{\Gamma}{1}{n[\tau]}{\kchildren[s_{l'}]_l}{n[\tau']}{L[l' {\To} l]}}
    {\wfupdeff{\Gamma}{*}{\tau}{s_{l'}}{\tau'}{L}}
    \quad
    \infer{\wfupdeff{\Gamma}{a}{\tau}{\kleft[s_{l'}]_l}{\tau',\tau}{L[l'{\To}l]}}
    {\wfupdeff{\Gamma}{*}{\emptyseq}{s}{\tau'}{L}}
    \quad
    \infer{\wfupdeff{\Gamma}{a}{\tau}{\kright[s_{l'}]_l}{\tau,\tau'}{L[l'{\To}l]}}
    {\wfupdeff{\Gamma}{*}{\emptyseq}{s}{\tau'}{L}}
  \end{array}\]
  \fbox{$\wfitereff{\Gamma}{\tau}{s}{\tau'}{L}$}
  \[\small\begin{array}{c}
    \infer{\wfitereff{\Gamma}{\emptyseq}{s_{l}}{\emptyseq}{\{l\}}}{}
    \quad
    \infer{\wfitereff{\Gamma}{\alpha}{s}{\tau}{L}}
    {\wfupdeff{\Gamma}{1}{\alpha}{s}{\tau}{L}}
    \quad
    \infer{\wfitereff{\Gamma}{\tau_1^*}{s}{\tau_2^*}{L}}
    {\wfitereff{\Gamma}{\tau_1}{s}{\tau_2}{L}}
\quad
    \infer{\wfitereff{\Gamma}{X}{s}{\tau}{L}}
    {\wfitereff{\Gamma}{E(X)}{s}{\tau}{L}}
    \smallskip\\
    \infer{\wfitereff{\Gamma}{\tau_1,\tau_2}{s}{\tau_1',\tau_2'}{L_1 \cap L_2}}
    {\wfitereff{\Gamma}{\tau_1}{s}{\tau_1'}{L_1}
      &
      \wfitereff{\Gamma}{\tau_2}{s}{\tau_2'}{L_2}}
    \quad
    \infer{\wfitereff{\Gamma}{\tau_1|\tau_2}{s}{\tau_1'|\tau_2'}{L_1 \cap L_2}}
    {\wfitereff{\Gamma}{\tau_1}{s}{\tau_1'}{L_1}
      &
      \wfitereff{\Gamma}{\tau_2}{s}{\tau_2'}{L_2}}
  \end{array}\]
  \caption{Path-error analysis for updates}\labelFig{path-analysis}
\end{figure*}

Besides developing a type system for \muXQ, \citet{colazzo06jfp}
studied the problem of identifying subexpressions of the query that
always evaluate to $\emptyseq$, but are not syntactically equal to
$\emptyseq$.  Such ``unproductive'' subexpressions typically indicate
errors in a query.  For example, the query $\forreturn{y\in x/a}{a[]}$
is well-formed in context $\Gamma = x:b[c[]^*,d[]^*]$, but
unproductive when evaluated against $\Gamma$ since $x/a$ will always
be empty.  \citet{colazzo06jfp} formally defined such
\emph{path-errors}\footnote{Arguably, the term ``path-errors'' is
  inaccurate in that there are expressions such as
  $\forreturn{\bar{x} \in ()}{\bar{x}}$ that do not mention path expressions, yet
  contain path-errors.  Nevertheless, we follow the existing
  terminology here.} and introduced a type-based analysis that detects
them.  In this section, we define path-errors for updates and derive a
path-error analysis for core \Flux.  We first introduce technical
machinery, then define path-errors and the analysis, and prove its
correctness.

Consider \emph{locations} $l$.  We will work with \emph{distinctly
  labeled statements} $s_l$ in which each core \Flux subexpression
carries a distinct location $l$.  We ignore locations as convenient
when we wish to view $s_l$ as an ordinary statement $s$.  Suppose $s$
is distinctly labeled.  We write $s[l]$ for the unique subexpression
of $e$ labeled by $l$ and write $s|_l$ for the result of replacing the
subexpression at $l$ in $s$ with $\kskip$.  For example,
$(\kiter[\kchildren[s_l]_{l'}]_{l''})|_{l'} = \kiter[\kskip]$.

We now define a form of path-errors suitable for updates, based on
replacing subexpressions with the trivial update $\kskip$ instead of
the empty sequence $\emptyseq$.
\begin{definition}
  Suppose $\wfupd{\Gamma}{a}{\tau}{s}{\tau'}$, where $s$ is distinctly
  labeled.  We say $s$ is \emph{unproductive} at $l$ provided
  $\evalu{\gamma}{v}{s}{v'} \iff \evalu{\gamma}{v}{s|_l}{v'}$ for
  every $\gamma \in \SB{\Gamma}, v \in \SB{\tau}, v' \in \SB{\tau'}$.
  Recall that update evaluation is functional so this means that $s$
  and $s|_l$ are equivalent over inputs from  $\SB{\Gamma},\SB{\tau}$.

  Moreover, we say that $s$ has an \emph{update path-error} at $l$
  provided $s$ is unproductive at $l$ and $s[l] \neq \kskip$, and say
  that $s$ is \emph{update path-correct} if $s$ has no update path
  errors.
\end{definition}

We define a static analysis for identifying update path-errors via the
rules in \refFig{path-analysis}.  The main judgment is
$\wfupdeff{\Gamma}{a}{\tau}{s_l}{\tau'}{L}$, meaning \emph{$s$ is
  well-formed and is unproductive at each $l \in L$}.  We employ an
auxiliary judgment $\wfitereff{\Gamma}{\tau}{s_l}{\tau'}{L}$ to handle
iteration.  We also define a ``conditional union'' operation:
\[
L[l_1,\ldots,l_n \To l] = \left\{
\begin{array}{ll} L \cup \{l\} & \{l_1,\ldots,l_n\} \subseteq L \\
L & \text{otherwise}
\end{array}\right.
\] 

Note that the analysis is intraprocedural.  It gives up when we
consider a procedure call $P(\vec{e})$: we conservatively assume that
there is no path error at $P(\vec{e})$, and we do not proceed to
analyze the body of $P$.  We can, of course, extend the analysis to
declarations $\Delta$ by analyzing each procedure body individually.

We first establish that the analysis produces results whenever $s$ is
well-formed.  This is straightforward by induction on derivations.
\begin{lemma}
~
\begin{enumerate}
  \item If $\wfupd{\Gamma}{a}{\tau}{s}{\tau'}$ then there exists $L$ such
  that $\wfupdeff{\Gamma}{a}{\tau}{s}{\tau'}{L}$.
\item If $\wfiter{\Gamma}{\tau}{s}{\tau'}$ then there exists $L$ such
  that $\wfitereff{\Gamma}{\tau}{s}{\tau'}{L}$.
\end{enumerate}\end{lemma}

The goal of the analysis is to conservatively \emph{underestimate} the
set of possible unproductive locations in $s$.
\begin{theorem}[Path-Error Analysis Soundness]
~
\begin{enumerate}
\item If $\wfupdeff{\Gamma}{a}{\tau}{s}{\tau'}{L}$ then $s$ is
  unproductive at every $l \in L$.
\item If $\wfitereff{\Gamma}{\tau}{s}{\tau'}{L}$ then $\kiter[s]$ is
  unproductive at every $l \in L$.
\end{enumerate}
\end{theorem}
Moreover, all of the labels in the set $\{l \in L\mid
s[l] \neq \kskip\}$ can be reported as update path-errors and used
to optimize $s$ by replacing each $s[l]$ with $\kskip$.

Using more sophisticated rules for typechecking $\klet$- and
$\kfor$-expressions, \cite{colazzo06jfp} were also able to show that
their path-error analysis is \emph{complete} for \muXQ (without
recursion); thus, path-correctness is decidable for the fragment of
XQuery they studied --- a nontrivial result.  Similar techniques can
be used to make update path-error analysis more precise, but it is not
obvious that this yields a complete analysis, even in the absence of
recursion.  We leave this issue for future work.

It is, of course, also of interest to perform path-error analysis at
the source level, so that the errors can be reported in terms familiar
to the user.  We believe that the path-error analysis can be
``lifted'' to the source type system, but leave this for future work.
However, it appears that many path-errors show up in the source type
system as empty substitutions $\Theta$ resulting from analyzing path
expressions.

\section{Related work}\labelSec{related}

\paragraph{Other database update languages}

\citet{DBLP:conf/ssdbm/LiefkeD99}'s update language CPL+, a typed
language for updating complex-object databases using path-based
insert, update, and delete operations.  High-level CPL+ updates were
translated to a simpler core language with orthogonal operations for
iteration, navigation, insertion/deletion, and replacement.  The \Flux
core language was strongly influenced by CPL+.

\paragraph{Static typing for XML processing}

We will focus on only the most closely related work on XML
typechecking; \citet{DBLP:conf/icdt/MollerS05} provide a much more
complete survey of type systems for XML transformation languages.

\citet{hosoya05toplas} introduced XDuce, the first statically typed
XML transformation language based on regular expressions.  
\citet{fernandez01icdt} introduced many of
the ideas for using XDuce-style regular expression subtyping for
typechecking an XML query language based on monadic comprehensions.
XQuery's type system~\cite{xquery-semantics-w3c-20070123} is also
based on regular expression types and subtyping, but its rules for
typechecking iteration are relatively imprecise: they discard
information about the order and multiplicity of the elements of a
sequence.  As discussed by \citet{cheney08esop}, taking this approach
to typechecking iterations in \emph{updates} would be disastrous since
many updates iterate over a part of the database while leaving its
structure intact.  \citet{colazzo06jfp} showed how to provide more
precise regular expression types to XQuery $\kfor$-iteration; we have
already discussed this work in the body of the paper.
\citet{cheney08esop} showed how to add subtyping and subsumption to
\muXQ and \Flux while retaining decidable typechecking.

\paragraph{XML update languages}

\cite{DBLP:conf/planX/Cheney07} provided a detailed discussion of XML update
language proposals and compared them with the \Flux approach.  Here,
we will only discuss closely related or more recent work.

\citet{calcagno05popl} investigated DOM-style XML updates using
\emph{context logic}, a logic of ``trees with holes''.
\citet{calcagno05popl} studied a Hoare-style logic for sequences of
atomic update operations on unordered XML.  \citet{gardner08pods}
extended this approach to ordered XML and while-programs over atomic
DOM updates.  This approach is very promising for reasoning about
low-level DOM updates, for example in Java or JavaScript programs.  It
should be possible to translate core \Flux to their variant of DOM; it
would interesting to see whether \Flux type information can also be
compiled down to context logic in an appropriate way.

The W3C XQuery Update Facility \cite{xquery-update-w3c-10-20080314} has
been under development for several years.  However, the typing rules in the
current draft treat updates as expressions of type $()$, and to
our knowledge this type system has not been proved sound.

\citet{ghelli07dbpl} have developed XQueryU, a variant of \XQueryBang
that is translated to an ``algebraic'' core language intended for
optimization.  However, the semantics of the core language is
defined by translation back to XQueryU, which seems circular.

Static typechecking has not been studied for any other extant XML
update language proposals, even though the W3C's XQuery Update
Facility Requirements
document~\cite{xquery-update-requirements-w3c-062006} lists static
typechecking as a strong requirement.  \Flux shows that applying
well-known functional language design principles leads to a language
with a relatively simple semantics and relatively straightforward type
system.

Static analysis techniques have been studied for only a few of these
languages.  \citet{DBLP:conf/cav/BenediktBFV05} and
\citet{DBLP:conf/ximep/BenediktBFV05} studied static analysis
techniques for optimizing updates in UpdateX, an earlier XML update
language proposal due to \citet{sur04planx}.  \citet{ghelli07icdt}
have developed a commutativity analysis for determining when two
side-effecting expressions in \XQueryBang can be reordered.
No prior work has addressed path-error or dead
code analysis for XML updates.

The design goals of many of these
proposals differ from those that motivate this work.  \Flux is not
meant to be a full-fledged programming language for mutable XML data.
Instead, it is meant to play a role for XML and XQuery similar to that
of SQL's update facilities relative to relational databases and SQL.
Its goal is only to be expressive enough for typical updates to XML
databases while remaining simple and statically typecheckable.

\paragraph{Mutability in functional languages} \Flux takes a ``purely
functional'' approach to typechecking updates.  The type of an update
reflects the changes to the mutable store an update may make.  This is
similar to side-effect encapsulation using monads or arrows in
Haskell.  An alternative possibility might be to use ML-like
references.  This could easily handle updates to parts of an XML
database whose type is fixed; however, handling updates that change
the \emph{type} of a part of the database would likely be problematic,
due to aliasing issues.  \Flux does not allow aliasing of the mutable
store, so avoids this problem.

\section{Extensions and future work}\labelSec{extensions}

\paragraph{Additional XQuery features}

To simplify the discussion, we have omitted features such as
attributes, comments, and processing instructions that are present in
the official XQuery data model, as well as the XPath axes needed to
access them.  We have also omitted the many additional base types and
built-in functions (such as \verb|position()| or \verb|last()|)
present in full XQuery/XPath.  All of these features can be added
without damaging the formal properties of the core language.

We have also omitted the descendant axis.  Many DTDs and XML Schemas
encountered in database applications of XML are nonrecursive and
``shallow'' \cite{DBLP:conf/webdb/Choi02,bex04webdb}.  Thus, in
practice, vertical recursion (the descendant axis \verb|//|) can
usually be avoided.  Simple updates involving \verb|//| can, however,
be simulated using recursive update procedures; in fact, for
non-recursive input types, updates involving \verb|//| can often
already be expressed in \Flux.  Further work is needed to
understand the expressiveness and usability tradeoffs involved in
typechecking more complex updates involving recursive types.

\paragraph{Transformations}

The XQuery Update Facility \cite{xquery-update-w3c-10-20080314}
includes \emph{transformations}, which allow running an update
operation within an XQuery expression, with side-effects confined
(somewhat like \verb|runST| in Haskell).  Such a facility can easily
be added using \Flux updates:
\[e ::= \cdots \mid \transform{e}{s}\] with semantics and typing
rules:
\[
\infer{\eval{\gamma}{\transform{e}{s}}{v'}}{\eval{\gamma}{e}{v} &
  \evalu{\gamma}{v}{s}{v'}} \quad
\infer{\wf{\Gamma}{\transform{e}{s}}{\tau_2}}{\wf{\Gamma}{e}{\tau_1} &
  \wfupd{\Gamma}{*}{\tau_1}{s}{\tau_2}}
\]

\paragraph{Dynamic typechecking, incremental validation and maintenance}

We believe it is important to combine \Flux-style static typechecking
with efficient dynamic techniques in order to handle cases where
static type information is imprecise.
\citet{DBLP:conf/icde/BarbosaMLMA04} and
\citet{DBLP:journals/tods/BalminPV04} have studied efficient
\emph{incremental validation} techniques for checking that sequences
of atomic updates preserve a database's schema.  These techniques
impose (manageable, but nonzero) run-time costs per atomic update
operation and storage overhead proportional to the database size;
also, they require that the input and output types are equal, a
significant limitation compared to \Flux.

\paragraph{Efficient implementation within XML databases}

We have built a prototype \Flux interpreter in OCaml, in order to
validate our type system and normalization translation designs and
experiment with variations.  The obvious next step is developing
efficient implementations of \Flux, particularly within XML database
systems.  \citet{DBLP:conf/ssdbm/LiefkeD99} investigated efficient
implementation techniques for CPL+ updates to complex-object
databases, which have much in common with XML databases.  One initial
implementation strategy could simply be to generate \XQueryBang or
XQuery Updates from core \Flux after normalization, typechecking and
high-level optimization; this should not be difficult since these
languages are more expressive than \Flux.  However, more sophisticated
techniques may be necessary to obtain good performance.

\section{Conclusions}\labelSec{concl}

The problem of updating XML databases poses many challenges to
language design. In previous work, we introduced \Flux, a simple core
language for XML updates, inspired in large part by the language CPL+
introduced by \citet{DBLP:conf/ssdbm/LiefkeD99} for updating complex
object databases.  In contrast to all other update proposals of which
we are aware, \Flux preserves the good features of XQuery such as its
purely functional semantics, while offering features convenient for
updating XML.  Moreover, \Flux is the first proposal for updating XML
to be equipped with a sound, static type system.

In this paper we have further developed the foundations of \Flux,
relaxing the limitations present in our preliminary proposal.  First,
we have extended its operational semantics and type system to handle
\emph{recursive types and updates}.  This turned out to be
straightforward.  Second, although the \Flux core language is easy to
understand, typecheck and optimize, it is not easy to use.  Therefore,
we have developed a \emph{high-level source language} for updates, and
shown how to translate it to core \Flux.  Since it is difficult to
propagate useful type error information from translated updates back
to source updates, we have also developed a type system for the source
language, and validated its design by proving that the translation
both \emph{preserves} and \emph{reflects} typability.  Third, we
developed a novel definition of \emph{update path-errors}, a form of
dead code analysis, and introduced a static analysis that identifies
them.

At present we have implemented a proof-of-concept prototype \Flux
interpreter, including typechecking for the source language and core
language, normalization, and path-error analysis. There are many
possible directions for future work; the most immediate is to develop
efficient optimizing implementations of \Flux within existing XML
databases or other XML-processing systems.

\small
\bibliographystyle{plainnat}
\bibliography{xupdate,xml,db,fp}

\begin{thebibliography}{32}
\providecommand{\natexlab}[1]{#1}
\providecommand{\url}[1]{\texttt{#1}}
\expandafter\ifx\csname urlstyle\endcsname\relax
  \providecommand{\doi}[1]{doi: #1}\else
  \providecommand{\doi}{doi: \begingroup \urlstyle{rm}\Url}\fi

\bibitem[Balmin et~al.(2004)Balmin, Papakonstantinou, and
  Vianu]{DBLP:journals/tods/BalminPV04}
A.~Balmin, Y.~Papakonstantinou, and V.~Vianu.
\newblock Incremental validation of {XML} documents.
\newblock \emph{ACM Transactions on Database Systems}, 29\penalty0
  (4):\penalty0 710--751, 2004.

\bibitem[Barbosa et~al.(2004)Barbosa, Mendelzon, Libkin, Mignet, and
  Arenas]{DBLP:conf/icde/BarbosaMLMA04}
D.~Barbosa, A.~O. Mendelzon, L.~Libkin, L.~Mignet, and M.~Arenas.
\newblock Efficient incremental validation of {XML} documents.
\newblock In \emph{ICDE}, pages 671--682. IEEE Computer Society, 2004.

\bibitem[Benedikt et~al.(2005{\natexlab{a}})Benedikt, Bonifati, Flesca, and
  Vyas]{DBLP:conf/cav/BenediktBFV05}
Michael Benedikt, Angela Bonifati, Sergio Flesca, and Avinash Vyas.
\newblock Verification of tree updates for optimization.
\newblock In Kousha Etessami and Sriram~K. Rajamani, editors, \emph{CAV},
  volume 3576 of \emph{Lecture Notes in Computer Science}, pages 379--393.
  Springer, 2005{\natexlab{a}}.

\bibitem[Benedikt et~al.(2005{\natexlab{b}})Benedikt, Bonifati, Flesca, and
  Vyas]{DBLP:conf/ximep/BenediktBFV05}
Michael Benedikt, Angela Bonifati, Sergio Flesca, and Avinash Vyas.
\newblock Adding updates to {XQuery}: Semantics, optimization, and static
  analysis.
\newblock In Daniela Florescu and Hamid Pirahesh, editors, \emph{XIME-P},
  2005{\natexlab{b}}.

\bibitem[Benzaken et~al.(2003)Benzaken, Castagna, and Frisch]{benzaken03icfp}
V\'eronique Benzaken, Giuseppe Castagna, and Alain Frisch.
\newblock {CDuce}: an {XML}-centric general-purpose language.
\newblock In \emph{ICFP '03: Proceedings of the eighth ACM SIGPLAN
  international conference on Functional programming}, pages 51--63, New York,
  NY, USA, 2003. ACM Press.

\bibitem[Bex et~al.(2004)Bex, Neven, and den Bussche]{bex04webdb}
Geert~Jan Bex, Frank Neven, and Jan~Van den Bussche.
\newblock {DTDs} versus {XML} schema: a practical study.
\newblock In \emph{WebDB '04: Proceedings of the 7th International Workshop on
  the Web and Databases}, pages 79--84, New York, NY, USA, 2004. ACM Press.

\bibitem[Calcagno et~al.(2005)Calcagno, Gardner, and Zarfaty]{calcagno05popl}
Cristiano Calcagno, Philippa Gardner, and Uri Zarfaty.
\newblock Context logic and tree update.
\newblock In \emph{Proceedings of the 32nd ACM SIGPLAN-SIGACT Symposium on
  Principles of Programming Languages (POPL 2005)}, pages 271--282, New York,
  NY, USA, 2005. ACM.

\bibitem[Chamberlin and Robie(2005)]{xquery-update-requirements-w3c-062006}
Don Chamberlin and Jonathan Robie.
\newblock {XQuery} update facility requirements.
\newblock W3C Working Draft, June 2005.
\newblock
  \texttt{http://www.w3.org/}\-\texttt{TR/}\-\texttt{xquery-update-requirement%
s/}.

\bibitem[Chamberlin et~al.(2006)Chamberlin, Carey, Florescu, Kossmann, and
  Robie]{chamberlin06ximep}
Don Chamberlin, Mike Carey, Daniela Florescu, Donald Kossmann, and Jonathan
  Robie.
\newblock {XQueryP}: Programming with {XQuery}.
\newblock In \emph{XIME-P}, 2006.

\bibitem[Chamberlin et~al.(2008)Chamberlin, Florescu, Melton, Robie, and
  Sim\'eon]{xquery-update-w3c-10-20080314}
Don Chamberlin, Daniela Florescu, Jim Melton, Jonathan Robie, and J\'er\^ome
  Sim\'eon.
\newblock {XQuery} update facility.
\newblock W3C Candidate Recommendation, March 2008.
\newblock
  \texttt{http://www.w3c.org/}\-\texttt{TR/}\-\texttt{xquery-update-10/}.

\bibitem[Cheney(2007)]{DBLP:conf/planX/Cheney07}
James Cheney.
\newblock Lux: A lightweight, statically typed {XML} update language.
\newblock In \emph{PLAN-X}, pages 25--36, 2007.

\bibitem[Cheney(2008)]{cheney08esop}
James Cheney.
\newblock Regular expression subtyping for {XML} query and update languages.
\newblock In \emph{Proceedings of the 17th European Symposium on Programming
  (ESOP 2008)}, number 4960 in LNCS, pages 32--46, 2008.

\bibitem[Choi(2002)]{DBLP:conf/webdb/Choi02}
Byron Choi.
\newblock What are real {DTDs} like?
\newblock In \emph{WebDB}, pages 43--48, 2002.

\bibitem[Clark(1999)]{clark99xslt}
J.~Clark.
\newblock {XSL} transformations {(XSLT)}.
\newblock W3C Recommendation, November 1999.
\newblock \verb|http://www.w3.org/TR/xslt|.

\bibitem[Colazzo et~al.(2006)Colazzo, Ghelli, Manghi, and
  Sartiani]{colazzo06jfp}
Dario Colazzo, Giorgio Ghelli, Paolo Manghi, and Carlo Sartiani.
\newblock Static analysis for path correctness of {XML} queries.
\newblock \emph{J. Funct. Program.}, 16\penalty0 (4-5):\penalty0 621--661,
  2006.

\bibitem[Draper et~al.(2007)Draper, Fankhauser, Fern\'andez, Malhotra, Rose,
  Rys, Sim\'eon, and Wadler]{xquery-semantics-w3c-20070123}
Denise Draper, Peter Fankhauser, Mary Fern\'andez, Ashok Malhotra, Kristoffer
  Rose, Michael Rys, J\'er\^ome Sim\'eon, and Philip Wadler.
\newblock {XQuery} 1.0 and {XPath} 2.0 formal semantics.
\newblock W3C Recommendation, January 2007.
\newblock \verb|http://www.w3.org/TR/xquery-semantics/|.

\bibitem[Fernandez et~al.(2001)Fernandez, Sim\'eon, and
  Wadler]{fernandez01icdt}
Mary~F. Fernandez, J\'er\^ome Sim\'eon, and Philip Wadler.
\newblock A semi-monad for semi-structured data.
\newblock In \emph{Proceedings of the 8th International Conference on Database
  Theory (ICDT 2001)}, pages 263--300, London, UK, 2001. Springer-Verlag.

\bibitem[Frisch(2006)]{frisch06icfp}
Alain Frisch.
\newblock {OCaml} + {XDuce}.
\newblock In \emph{ICFP '06: Proceedings of the eleventh ACM SIGPLAN
  international conference on Functional programming}, pages 192--200, New
  York, NY, USA, 2006. ACM Press.

\bibitem[Gapeyev et~al.(2006)Gapeyev, Garillot, and
  Pierce]{DBLP:conf/planX/GapeyevGP06}
Vladimir Gapeyev, Fran\c{c}ois Garillot, and Benjamin~C. Pierce.
\newblock Statically typed document transformation: An {Xtatic} experience.
\newblock In Giuseppe Castagna and Mukund Raghavachari, editors, \emph{PLAN-X},
  pages 2--13. BRICS, 2006.

\bibitem[Gardner et~al.(2008)Gardner, Smith, Wheelhouse, and
  Zarfaty]{gardner08pods}
Philippa Gardner, Gareth Smith, Mark Wheelhouse, and Uri Zarfaty.
\newblock Local hoare reasoning about {DOM}.
\newblock In \emph{Proceedings of the 2008 Symposium on Principles of Database
  Systems (PODS 2008)}, pages 261--270, 2008.

\bibitem[Ghelli et~al.(2007{\natexlab{a}})Ghelli, Rose, and
  Sim\'eon]{ghelli07icdt}
G.~Ghelli, K.~Rose, and J.~Sim\'eon.
\newblock Commutativity analysis in {XML} update languages.
\newblock In Dan Suciu and Thomas Schwentick, editors, \emph{Proceedings of the
  11th International Conference on Database Theory (ICDT 2007)}, pages
  374--388, January 2007{\natexlab{a}}.

\bibitem[Ghelli et~al.(2006)Ghelli, Re, and
  Sim{\'e}on]{DBLP:conf/edbtw/GhelliRS06}
Giorgio Ghelli, Christopher Re, and J{\'e}r{\^o}me Sim{\'e}on.
\newblock {XQuery!}: An {XML} query language with side effects.
\newblock In \emph{EDBT Workshops}, volume 4254 of \emph{Lecture Notes in
  Computer Science}, pages 178--191. Springer, 2006.

\bibitem[Ghelli et~al.(2007{\natexlab{b}})Ghelli, Onose, Rose, and
  Sim{\'e}on]{ghelli07dbpl}
Giorgio Ghelli, Nicola Onose, Kristoffer Rose, and J{\'e}r{\^o}me Sim{\'e}on.
\newblock A better semantics for {XQuery} with side-effects.
\newblock In Marcelo Arenas and Michael~I. Schwartzbach, editors, \emph{DBPL},
  volume 4797 of \emph{Lecture Notes in Computer Science}, pages 81--96.
  Springer, 2007{\natexlab{b}}.

\bibitem[Hosoya and Pierce(2003)]{hosoya03toit}
Haruo Hosoya and Benjamin~C. Pierce.
\newblock {XDuce}: A statically typed {XML} processing language.
\newblock \emph{ACM Trans. Internet Technology}, 3\penalty0 (2):\penalty0
  117--148, 2003.

\bibitem[Hosoya et~al.(2005)Hosoya, Vouillon, and Pierce]{hosoya05toplas}
Haruo Hosoya, J\'er\^ome Vouillon, and Benjamin~C. Pierce.
\newblock Regular expression types for {XML}.
\newblock \emph{ACM Trans. Program. Lang. Syst.}, 27\penalty0 (1):\penalty0
  46--90, 2005.

\bibitem[Hughes(2000)]{hughes00scp}
John Hughes.
\newblock Generalising monads to arrows.
\newblock \emph{Sci. Comput. Program.}, 37\penalty0 (1-3):\penalty0 67--111,
  2000.

\bibitem[Laux and Martin(2000)]{laux00xmldb}
Andreas Laux and Lars Martin.
\newblock {XUpdate} - {XML} update language.
\newblock
  \texttt{http://xmldb-org.sourceforge.net/}\-\texttt{xupdate/}\-\texttt{xupda%
te-wd.html}, September 2000.
\newblock Work in progress.

\bibitem[Liefke and Davidson(1999)]{DBLP:conf/ssdbm/LiefkeD99}
Hartmut Liefke and Susan~B. Davidson.
\newblock Specifying updates in biomedical databases.
\newblock In \emph{SSDBM}, pages 44--53, 1999.

\bibitem[M{\o}ller and Schwartzbach(2005)]{DBLP:conf/icdt/MollerS05}
Anders M{\o}ller and Michael~I. Schwartzbach.
\newblock The design space of type checkers for {XML} transformation languages.
\newblock In \emph{Proceedings of the 10th International Conference on Database
  Theory (ICDT 2005)}, pages 17--36, 2005.

\bibitem[Sur et~al.(2004)Sur, Hammer, and Sim\'eon]{sur04planx}
Gargi Sur, Joachim Hammer, and J\'er\^ome Sim\'eon.
\newblock {UpdateX} - an {XQuery}-based language for processing updates in
  {XML}.
\newblock In \emph{PLAN-X}, 2004.

\bibitem[Tatarinov et~al.(2001)Tatarinov, Ives, Halevy, and
  Weld]{DBLP:conf/sigmod/TatarinovIHW01}
Igor Tatarinov, Zachary~G. Ives, Alon~Y. Halevy, and Daniel~S. Weld.
\newblock Updating {XML}.
\newblock In \emph{SIGMOD Conference}, pages 413--424, 2001.

\bibitem[Wang et~al.(2003)Wang, Liu, and Lu]{DBLP:conf/ideas/WangLL03}
Guoren Wang, Mengchi Liu, and Li~Lu.
\newblock Extending {XML-RL} with update.
\newblock In \emph{IDEAS}, pages 66--75. IEEE Computer Society, 2003.

\end{thebibliography}

\newpage
\normalsize
\appendix

\section{Semantics and type system for \muXQ queries}

We will use an XQuery-like core language called \muXQ, introduced by
Colazzo et al. (2006).  Following that paper, we distinguish between
\emph{tree variables} $\bar{x} \in \TVar$, introduced by $\kfor$, and
\emph{forest variables}, $x \in \Var$, introduced by $\klet$.  The
other syntactic classes of our variant of \muXQ include labels $l,m,n
\in Lab$ and expressions $e \in \Expr$; the abstract syntax of
expressions is defined by the following BNF grammar:
\begin{eqnarray*}
  e &::=& \emptyseq \mid e,e' \mid n[e] \mid w \mid x \mid \letin{x=e}{e'}\\
  &\mid& \ktrue \mid \kfalse\mid \ifthenelse{c}{e}{e'} \mid e \eq e'\\
  &\mid & \bar{x} \mid \bar{x}/\kchild \mid e::n  \mid \forreturn{\bar{x} \in e}{e'}
\end{eqnarray*}
The distinguished variables $\bar{x}$ in $\forreturn{\bar{x} \in
  e}{e'(\bar{x})}$ and $x$ in $\letin{x=e}{e''(x)}$ are bound in
$e'(\bar{x})$ and $e''(x)$ respectively.  Here and elsewhere, we
employ common conventions such as considering expressions containing
bound variables equivalent up to $\alpha$-renaming and employing a
richer concrete syntax including, for example, parentheses.

Recursive queries can be added to \muXQ in the same manner as in
XQuery without damaging the properties of the system needed in this
paper.

To simplify the presentation, we split \muXQ's projection operation
$\bar{x}~\kchild::l$ into two expressions: child projection
($\bar{x}/\kchild$) which returns the children of $\bar{x}$, and node
name filtering ($e::n$) which evaluates $e$ to an arbitrary sequence
and selects the nodes labeled $n$.  Thus, the ordinary child axis
expression $\bar{x}~\kchild::n$ is syntactic sugar for $
(\bar{x}/\kchild)::n$ and the ``wildcard'' child axis is definable as
$\bar{x}~\kchild::* = \bar{x}/\kchild$.  We also consider only one
built-in operation, string equality.

\begin{figure}
  \begin{eqnarray*}
    \ichildren(n[f]) &=& f\\
    \ichildren(v) &=& \emptyseq \quad (v \not\eq n[v'])
  \end{eqnarray*}
  \[\begin{array}{rclcrcl}
    \SB{\ktrue}\gamma &=& \ktrue&&
    \SB{\kfalse}\gamma &=& \kfalse\\
    \SB{\emptyseq}\gamma &=& \emptyseq&&
    \SB{e,e'}\gamma &=& \SB{e}\gamma, \SB{e'}\gamma\\
    \SB{n[e]}\gamma &=& n[\SB{e}\gamma]&&
    \SB{w}\gamma &=& w\\
    \SB{x}\gamma &=& \gamma(x)&&
    \SB{\bar{x}}\gamma &=& \gamma(\bar{x})\\
  \end{array}\]
  \begin{eqnarray*}
    \SB{\letin{x=e_1}{e_2}}\gamma &=& \SB{e_2}\gamma[x:=\SB{e_1}\gamma]\\
    \SB{\ifthenelse{c}{e_1}{e_2}}\gamma &=& 
    \left\{\begin{array}{ll}
        \SB{e_1}\gamma & \SB{c}\gamma \eq \ktrue\\
        \SB{e_2}\gamma & \SB{c}\gamma \eq \kfalse\\
      \end{array}\right.\\
    \SB{e = e'}\gamma &=& \left\{\begin{array}{ll} 
        \ktrue & \SB{e}\gamma \eq \SB{e'}\gamma
        \\
        \kfalse & \SB{e}\gamma \not\eq \SB{e'}\gamma
      \end{array}\right.\\
    \SB{e::n}\gamma &=& [n[v] \mid n[v] \in \SB{e}\gamma]\\
    \SB{\bar{x}/\kchild}\gamma &=& \ichildren(\gamma(\bar{x}))\\
    \SB{\forreturn{\bar{x} \in e_1}{e_2}}\gamma &=&[\SB{e_2}\gamma[\bar{x}:=t] \mid t \in \SB{e_1}\gamma]
  \end{eqnarray*}
  \caption{Semantics of query expressions.}\labelFig{query-semantics}
  \fbox{$\wf{\Gamma}{e}{\tau}$}
  \[
  \begin{array}{c}
    \infer{\wf{\Gamma}{\emptyseq}{\emptyseq}}{}
    \quad
    \infer{\wf{\Gamma}{w}{\kstring}}{}
    \smallskip\\
    \infer{\wf{\Gamma}{\ktrue}{\kbool}}{}
    \quad
    \infer{\wf{\Gamma}{\kfalse}{\kbool}}{}
    \smallskip\\
    \infer{\wf{\Gamma}{e,e'}{\tau,\tau'}}{\wf{\Gamma}{e}{\tau} & \wf{\Gamma}{e'}{\tau'}}
    \quad
    \infer{\wf{\Gamma}{n[e]}{n[\tau]}}{\wf{\Gamma}{e}{\tau}}
    \quad
    \infer{\wf{\Gamma}{e=e'}{\kbool}}{\wf{\Gamma}{e,e'}{\kstring}}
    \smallskip\\
    \infer{\wf{\Gamma}{\bar{x}}{\alpha}}{\bar{x}{:}\alpha \in \Gamma}
    \quad
    \infer{\wf{\Gamma}{x}{\tau}}{x{:}\tau \in \Gamma}
    \quad
    \infer{\wf{\Gamma}{\letin{x=e_1}{e_2}}{\tau_2}}
    {\wf{\Gamma}{e_1}{\tau_1} & \wf{\Gamma,x{:}\tau_1}{e_2}{\tau_2}}
    \smallskip\\
    \infer{\wf{\Gamma}{\ifthenelse{c}{e_1}{e_2}}{\tau_1|\tau_2}}
    {\wf{\Gamma}{c}{\kbool} & 
      \wf{\Gamma}{e_1}{\tau_1} &
      \wf{\Gamma}{e_2}{\tau_2} 
    }
    \smallskip\\
    \infer{\wf{\Gamma}{\bar{x}/\kchild}{\tau}}{\bar{x}{:}n[\tau] \in \Gamma}
    \quad
    \infer{\wf{\Gamma}{e::n}{\tau'}}{\wf{\Gamma}{e}{\tau} 
      & \tylab{\tau}{n}{\tau'}}
    \smallskip\\
    \infer{\wf{\Gamma}{\forreturn{\bar{x} \in e_1}{e_2}}{\tau_2}}
    {\wf{\Gamma}{e_1}{\tau_1} 
      & 
      \wfin{\Gamma}{x}{\tau_1}{e_2}{\tau_2}}
    \quad
    \infer{\wf{\Gamma}{e}{\tau'}}
    {\wf{\Gamma}{e}{\tau} & \tau \subty \tau'}
  \end{array}
  \]
  \fbox{$\tylab{\tau}{n}{\tau'}$}
  \[\begin{array}{c}
    \infer{\tylab{n[\tau]}{n}{n[\tau]}}{}
    \quad
    \infer{\tylab{\alpha}{n}{ \emptyseq}}{\alpha \neq n[\tau]}
    \quad 
    \infer{\tylab{\emptyseq}{n}{\emptyseq}}{}
    \quad
    \infer{\tylab{\tau_1^*}{n}{\tau_2^*}}
    {\tylab{\tau_1}{n}{\tau_2}}
    \smallskip\\
    \infer{\tylab{\tau_1,\tau_2}{n}{\tau_1',\tau_2'}}
    {\tylab{\tau_1}{n}{\tau_1'}
      &
      \tylab{\tau_2}{n}{\tau_2'}}
    \quad
    \infer{\tylab{\tau_1|\tau_2}{n}{\tau_1'|\tau_2'}}
    {\tylab{\tau_1}{n}{\tau_1'}
      &
      \tylab{\tau_2}{n}{\tau_2'}}
  \end{array}
  \]
  \fbox{$\wfin{\Gamma}{x}{\tau}{e}{\tau'}$}
  \[\begin{array}{c}
    \infer{\wfin{\Gamma}{x}{\emptyseq}{e}{\emptyseq}}{}
    \quad
    \infer{\wfin{\Gamma}{x}{\alpha}{e}{\tau}}{\wf{\Gamma,\bar{x}{:}\alpha}{e}{\tau}}
    \quad
    \infer{\wfin{\Gamma}{x}{\tau_1^*}{e}{\tau_2^*}}{\wfin{\Gamma}{x}{\tau_1}{e}{\tau_2}}
    \smallskip\\
    \infer{\wfin{\Gamma}{x}{\tau_1,\tau_2}{e}{\tau_1',\tau_2'}}
    {\wfin{\Gamma}{x}{\tau_1}{e}{\tau_1'} & 
      \wfin{\Gamma}{x}{\tau_2}{e}{\tau_2'}}
    \smallskip\\
    \infer{\wfin{\Gamma}{x}{\tau_1|\tau_2}{e}{\tau_1'|\tau_2'}}
    {\wfin{\Gamma}{x}{\tau_1}{e}{\tau_1'} & 
      \wfin{\Gamma}{x}{\tau_2}{e}{\tau_2'}}
  \end{array}\]
  \caption{Query well-formedness.}\labelFig{query-wf}
\end{figure}

\section{Type soundness for updates}

Type soundness for updates relies on pre-existing results for type
soundness for queries, which we repeat here:
\begin{theorem}[Query soundness]\labelThm{query-soundness-app}
  If $\wf{\Gamma}{e}{\tau}$ and $\gamma \in \SB{\Gamma}$ then
  $\eval{\gamma}{e}{v}$ with $v \in \SB{ \tau}$.
\end{theorem}

We need the following lemmas summarizing properties of test
subtyping and of the operational behavior of iteration.

\begin{lemma}[Subtyping and tests]\labelLem{subtyping-tests}
  If $t \in \SB{\alpha}$ then $\alpha \subty \phi$ if and only if
  $t \in \SB{\phi}$.
\end{lemma}
\begin{proof}
  First note that if $\tau \in \SB{\alpha} $ and $\alpha \subty \phi$
  then $t \in \SB{\phi}$.  For the reverse direction, suppose $\tau
  \in \SB{\alpha}$ and $\alpha \not\subty \phi$, and consider all
  combinations of cases.
\end{proof}

\begin{lemma}[Iteration]\labelLem{iteration}
  If $\evalu{\gamma}{v_1}{\kiter[s]}{v_1'}$,
  $\evalu{\gamma}{v_2}{\kiter[s]}{v_2'}$ are derivable then
  $\evalu{\gamma}{v_1,v_2}{\kiter[s]}{v_1',v_2'}$ is derivable.
\end{lemma}

\begin{theorem}[Update soundness]\labelThm{update-soundness-app}
  ~
  \begin{enumerate}
  \item If $\wfupd{\Gamma}{a}{\tau}{s}{\tau'}$, $v \in \SB{\tau}$, and
    $\gamma \in \SB{\Gamma}$, then $\evalu{\gamma}{v}{s}{v'}$ implies
    $v' \in \SB{ \tau'}$.
  \item If $\wfiter{\Gamma}{\tau}{s}{\tau'}$, $v \in \SB{\tau}$,
    and $\gamma \in \SB{\Gamma}$, then $\evalu{\gamma}{v}{\kiter[s]}{v'}$
    implies $v' \in \SB{ \tau'}$.
  \end{enumerate}
\end{theorem}
\begin{proof}
  Parts (1) and (2) must be proved simultaneously by induction; in
  each case the induction is on the typing derivation.  The cases
  involving standard constructs (if, let, skip, sequencing) are
  omitted.  For each case, we first show the typing derivation and
  then the (unique) corresponding operational derivation that can be
  constructed, with remarks as appropriate.
\begin{itemize}
\item Case ($\kinsert$): 
\[\infer{\wfupd{\Gamma}{*}{\emptyseq}{\kinsert~e}{\tau}}
{\wf{\Gamma}{e}{\tau}}
\quad\infer{\evalu{\gamma}{\emptyseq}{\kinsert~e}{v}}
{\eval{\gamma}{e}{v}}
\]
Follows by query soundness (\refThm{query-soundness}).
\item Case ($\kdelete$): 
\[\infer{\wfupd{\Gamma}{a}{\tau}{\kdelete}{\emptyseq}}
{}
\quad\infer{\evalu{\gamma}{v}{\kdelete}{\emptyseq}}
{}
\]
Immediate.
\item Case ($\krename$): 
\[\infer{\wfupd{\Gamma}{*}{n'[\tau]}{\krename~n}{n[\tau]}}
{}
\quad\infer{\evalu{\gamma}{n'[v]}{\krename~n}{n[v]}}
{}
\]
Follows since $n'[v] \in \SB{n'[\tau]}$ implies $v \in \SB{\tau}$ so
$n[v] \in \SB{n[\tau]}$.
\item Case ($\ksnapshot$): 
\[\infer{\wfupd{\Gamma}{a}{\tau}{\snapshot{x}{s}}{\tau'}}{\wfupd{\Gamma,x{:}\tau}{a}{\tau}{s}{\tau'}}
\quad\infer{\evalu{\gamma}{v}{\snapshot{x}{s}}{v'}}
{\evalu{\gamma[x:=v]}{v}{s}{v'}}
\]
Follows by induction, using the fact that $v \in \SB{\tau}$ so that
$\gamma[x:=v] \in \SB{\Gamma,x{:}\tau}$.
\item Case (test1): 
\[\infer{\wfupd{\Gamma}{1}{\alpha}{\phi?s}{\tau}}
{\alpha \subty \phi & \wfupd{\Gamma}{1}{\alpha}{s}{\tau}}
\quad
\infer{\evalu{\gamma}{t}{\phi?s}{v}}
{t \in \SB{\phi} & \evalu{\gamma}{t}{s}{v}}
\]
This case follows immediately by appealing to \refLem{subtyping-tests}
and then the induction hypothesis.
\item Case (test2): 
\[\infer{\wfupd{\Gamma}{1}{\alpha}{\phi?s}{\alpha}}
{\alpha \not\subty \phi}
\quad\infer{\evalu{\gamma}{t}{\phi?s}{t}}
{t \not\in \SB{\phi}}
\]
This case is immediate by \refLem{subtyping-tests}.  
\item Case (children): 
\[\infer{\wfupd{\Gamma}{1}{n[\tau]}{\kchildren[s]}{n[\tau']}}
{\wfupd{\Gamma}{*}{\tau}{s}{\tau'}}
\]
\[\infer{\evalu{\gamma}{n[v]}{\kchildren[s]}{n[v']}}
{\evalu{\gamma}{v}{s}{v'}}
\]
Clearly, since $n[v] \in \SB{n[\tau]}$, we must have $v \in \SB{\tau}$.  By
induction, we have that $v' \in \SB{\tau'}$, from which it is
immediate that $n[v'] \in \SB{n[\tau]}$.
\item Case ($\kright$): 
\[\infer{\wfupd{\Gamma}{a}{\tau}{\kright[s]}{\tau,\tau'}}
{\wfupd{\Gamma}{*}{\emptyseq}{s}{\tau'}}
\quad\infer{\evalu{\gamma}{v}{\kright[s]}{v,v'}}
{\evalu{\gamma}{\emptyseq}{s}{v'}}
\]
By assumption, $v \in \SB{\tau}$, induction, we have that $v' \in
\SB{\tau'}$, so $v,v' \in \SB{\tau,\tau'}$.
\item Case ($\kleft$): Symmetric.
\item Case ($\kiter$): 
\[\infer{\wfupd{\Gamma}{*}{\tau}{\kiter[s]}{\tau'}}
{\wfiter{\Gamma}{\tau}{s}{\tau'}}
\]
We proceed using induction hypothesis (3).
\item Case $P(\vec{e})$:
Suppose the typing derivation is of the form 
\[\infer{\wfupd{\Gamma}{a}{\sigma_1'}{P(\vec{e})}{\sigma_2}}{
\begin{array}{l}
  \procdecl{P(\vec{x}:\vec{\tau})}{\sigma_1}{\sigma_2}\defeq s \in \Delta \quad \sigma_1' \subty \sigma_1\\
\wf{\Gamma}{e_1}{\tau_1'} \quad \tau_1' \subty \tau_1\\
 \quad \cdots \quad\\
\wf{\Gamma}{e_n}{\tau_n'} \quad \tau_n' \subty \tau_n
\end{array}
}
\]
Hence the operational semantics derivation must be of the form:
\[\infer{\evalu{\gamma}{v}{P(\vec{e})}{v'}}{
\begin{array}{l}
\procdecl{P(\vec{x}:\vec{\tau})}{\sigma_1}{\sigma_2}\defeq s \in \Delta\\
\eval{\gamma}{e_1}{v_1} \\
\cdots\\
\eval{\gamma}{e_n}{v_n} \\
\evalu{\gamma[x_1:=v_1,\ldots,x_n:=v_n]}{v}{s}{v'}
\end{array}}\]
Then by query soundness and the definition of $\subty$ we have $v_i
\in \SB{\tau_i'} \subseteq\SB{\tau_i'}$ for each $i \in
\{1,\ldots,n\}$.  Hence, $\gamma[x_1:=v_1,\ldots,x_n:=v_n] \in
\SB{\Gamma,x_1{:}\tau_1,\ldots,x_n{:}\tau_n}$.  Moreover, $v \in
\SB{\sigma_1'}\subty \SB{\sigma_1}$ again by definition of subtyping,
so by induction, we can conclude that $v' \in \SB{\sigma_2}$ as
desired.

\item Case ($\kiter$1): If the derivation is of the form
\[\infer{\wfiter{\Gamma}{\emptyseq}{s}{\emptyseq}}{} \quad \infer{\evalu{\gamma}{\emptyseq}{\kiter[s]}{\emptyseq}}{}
\]
then the conclusion is immediate.

\item Case ($\kiter$2): If the derivation is of the form
\[\infer{\wfiter{\Gamma}{\alpha}{s}{\tau}}
{\wfupd{\Gamma}{1}{\alpha}{s}{\tau}}
\]
then by assumption, $v \in \SB{\alpha}$.  Hence, $v = t,\emptyseq$, so
by induction hypothesis (2), we have $\evalu{\gamma}{t}{s}{v'}$ with
$v' \in \SB{\tau}$ and can derive
\[\infer{\evalu{\gamma}{t,\emptyseq}{\kiter[s]}{v',\emptyseq}}
{\evalu{\gamma}{t}{s}{v'}
& 
\infer{\evalu{\gamma}{\emptyseq}{\kiter[s]}{\emptyseq}}{}}
\]

\item Case ($\kiter$3): If the derivation is of the form
\[\infer{\wfiter{\Gamma}{\tau_1,\tau_2}{s}{\tau_1',\tau_2'}}
{\wfiter{\Gamma}{\tau_1}{s}{\tau_1'}
&
\wfiter{\Gamma}{\tau_2}{s}{\tau_2'}}
\]
then we must have $v = v_1,v_2$ where $v_i \in \SB{\tau_i}$ for $i \in \{1,2\}$; by induction we have $\evalu{\gamma}{v_i}{\kiter[s]}{v_i'}$, where $v_i' \in \SB{\tau_i'}$ for $i \in \{1,2\}$.
Hence, by \refLem{iteration} we can conclude $\evalu{\gamma}{v_1,v_2}{\kiter[s]}{v_1',v_2'}$ where $v_1',v_2' \in \SB{\tau_1',\tau_2'}$.

\item Case ($\kiter$4): If the derivation is of the form
\[\infer{\wfiter{\Gamma}{\tau_1|\tau_2}{s}{\tau_1'|\tau_2'}}
{\wfiter{\Gamma}{\tau_1}{s}{\tau_1'}
&
\wfiter{\Gamma}{\tau_2}{s}{\tau_2'}}
\]
then we must have $v \in \SB{\tau_1} \cup \SB{\tau_2}$; the cases are symmetric, so without loss suppose $v \in \SB{\tau_1}$.  By
induction we have that $\evalu{\gamma}{v}{\kiter[s]}{v'}$ where $v' \in \SB{\tau_i'} \subty \SB{\tau_1 | \tau_2}$.

\item Case ($\kiter$5): If the derivation is of the form
\[\infer{\wfiter{\Gamma}{\tau_1^*}{s}{\tau_2^*}}
{\wfiter{\Gamma}{\tau_1}{s}{\tau_2}}
\]
then since $v \in \tau^*$ we must have that either $v = \emptyseq$ (in
which case the conclusion is immediate) or $v = v_1,\ldots,v_n$ where
each $v_i \in \SB{\tau}$.  By induction, we can obtain derivations
$\evalu{\gamma}{v_i}{\kiter[s]}{v_i'}$ where $v_i' \in \SB{\tau_2}$
for each $i \in\{1,\ldots,n\}$; hence, by repeated application of
\refLem{iteration} we can conclude that
$\evalu{\gamma}{v}{\kiter[s]}{v'}$, where by definition $v' =
v_1',\ldots,v_n' \in \SB{\tau_2^*}$.
\item Case ($\kiter$6): If the derivation is of the form
\[
    \infer{\wfiter{\Gamma}{X}{s}{\tau}}
    {\wfiter{\Gamma}{E(X)}{s}{\tau}}
\]
then we have that $v \in \SB{X} = \SB{E(X)}$ so the induction
hypothesis applies directly and we can conclude that $v' \in
\SB{\tau}$.
\end{itemize}
This exhausts all cases and completes the proof.
\end{proof}

\section{Normalizing and typechecking source updates}

Figures \ref{fig:source-tc-stmt}, \ref{fig:source-tc-upd},
\ref{fig:source-tc-path}, and \ref{fig:source-tc-filt} show the main
typechecking judgments for source statements, simple updates, and
paths.  The statement typechecking rules are straightforward; note
however that we require that both statements and updates start and end
with atomic types.  The simple update typechecking rules each follow
the procedure outlined above.  The path typechecking rules match $p$
against the input type $\alpha$.  Note that paths may bind variables,
and the same variable may be bound to different types for different
cases; this is why we need to include contexts $\Gamma$ in the
substitutions $\Theta$.  In certain rules, we choose fresh type
variables.  The scope with respect to which we require freshness is
all type variables mentioned elsewhere in the surrounding derivation.

For many of the typechecking judgments, we need to typecheck an
expression against all of the bindings of a context-tagged
substitution $\Theta$. We therefore introduce several
\emph{simultaneous typechecking} judgments shown in
\refFig{simult-source-tc}.

\subsection{Metatheory}

For this source language, we first need some auxiliary lemmas to establish soundness.
\begin{lemma}
  If $\wfupds{\Theta\oplus x}{s}{\Theta'\oplus x}$
  then $\wfupds{\Theta}{\snapshot{x}{s}}{\Theta'}$.
\end{lemma}
\begin{proof}
  Induction on derivation of $\wfupds{\Theta\oplus x}{s}{\Theta'\oplus x}$.
\end{proof}

We also need an auxiliary notation $\Theta|\Theta'$, which merges two context-tagged type substitutions provided their bindings and contexts match:
\[\small\begin{array}{rcl}
  \emptyset | \emptyset &=& \emptyset\\
(\Theta,Z \mapsto (\Gamma\tri \tau) ) | (\Theta',Z \mapsto (\Gamma\tri\tau')) &=& (\Theta | \Theta'), Z \mapsto (\Gamma,\tau|\tau')
\end{array}\]

\begin{lemma}
  If $\wfs{\Theta}{e}{\kbool}$ and $\wfupds{\Theta}{s}{\Theta'}$ then
  $\wfupds{\Theta}{\ifthenelse{e}{s}{\kskip}}{\Theta|\Theta'}$.
\end{lemma}
\begin{proof}
  Induction on derivation of $\wfupds{\Theta}{s}{\Theta'}$.
\end{proof}
\begin{lemma}
  \begin{enumerate}
  \item If the free type variables of $\tau$ are disjoint from those
    of $\Theta'$ then $\tau\subst{\Theta \uplus \Theta'} =
    \tau\subst{\Theta}$.
  \item If $\wfupdfilt{\Gamma}{\tau}{\phi}{\tau'}{\Theta}$ then
    $\tau=\tau'\subst{\Theta}$.
  \item If $\wfupdpath{\Gamma}{\alpha}{p}{\alpha'}{\Theta}$ then
    $\alpha = \alpha'\subst{\Theta}$.
  \item If $\wfupdpaths{\Theta}{p}{\Theta'}{\Theta''}$ then
    $\Theta = \Theta'\subst{\Theta''}$.
  \end{enumerate}
\end{lemma}
\begin{proof}
  For part (1), proof is by induction on the structure of types.  For
  part (2), proof is by induction on the structure of derivations,
  using part (1) for the cases involving types $\tau_1, \tau_2$ and
  $\tau_1 | \tau_2$.  Parts (3) and (4) follow by simultaneous
  induction on derivations.
\end{proof}
\begin{lemma}
  If $\wfupdfilt{\Gamma}{\tau}{\phi}{\tau'}{\Theta}$ and
  $\wfupds{\Theta}{s}{\Theta'}$ then
  $\wfiter{\Gamma}{\tau}{\phi?s}{\tau'\subst{\Theta'}}$.
\end{lemma}
\begin{proof}
  Induction on derivation of $\wfupdfilt{\Gamma}{\tau}{\phi}{\tau'}{\Theta}$.
\end{proof}
\begin{theorem}[Soundness]
  Assume $\Gamma,\alpha,\alpha',\Theta_1$ have no free type variables $Z$.
  ~\begin{enumerate}
\item If $\wfupdstmt{\Gamma}{\tau}{s}{\tau'}$ then
  $\wfupd{\Gamma}{1}{\tau}{\nstmt{s}}{\tau'}$.
\item If $\wfupdsimp{\Gamma}{\alpha}{u}{\alpha'}$ then 
  $\wfupd{\Gamma}{1}{\alpha}{\nupd{u}}{\alpha'}$.
\item If $\wfupdpath{\Gamma}{\alpha}{p}{\alpha'}{\Theta}$ and
  $\wfupds{\Theta}{s}{\Theta'}$ then
  $\wfupd{\Gamma}{1}{\alpha}{\npath{p}(s)}{\alpha'\subst{\Theta'}}$.
\item If $\wfupdpaths{\Theta_1}{p}{\Theta_2}{\Theta_3}$ and
  $\wfupds{\Theta_3}{s}{\Theta_3'}$ then
  $\wfupds{\Theta_1}{\npath{p}(s)}{\Theta_2\subst{\Theta_3'}}$.
\end{enumerate}
\end{theorem}

Now, to prove completeness, we need lemmas establishing that the earlier lemmas are invertible:
\begin{lemma}
  If $\wfupds{\Theta}{\snapshot{x}{s}}{\Theta'}$ then
  $\wfupds{\Theta\oplus x}{s}{\Theta'\oplus x}$.
\end{lemma}
\begin{proof}
  Induction on the structure of derivations of
  $$\wfupds{\Theta}{\snapshot{x}{s}}{\Theta'}$$ followed by inversion.
\end{proof}
\begin{lemma}
  If $\wfupds{\Theta}{\ifthenelse{e}{s}{\kskip}}{\Theta'}$ then there exists
  $\Theta''$ such that $\Theta' = \Theta|\Theta''$ and
  $\wfs{\Theta}{e}{\kbool}$ and $\wfupds{\Theta}{s}{\Theta''}$.
\end{lemma}
\begin{proof}
  Induction on the structure of derivations of
  $$\wfupds{\Theta}{\ifthenelse{e}{s}{\kskip}}{\Theta'}$$ followed by inversion.
\end{proof}
\begin{lemma}
  If $\wfiter{\Gamma}{\tau}{\phi?s}{\tau'}$ then there exists $\tau'',
  \Theta,\Theta'$ such that $\tau''\subst{\Theta'} = \tau'$, 
  $\wfupdfilt{\Gamma}{\tau}{\phi}{\tau''}{\Theta}$ and
  $\wfupds{\Theta}{s}{\Theta'}$.
\end{lemma}
\begin{proof}
  Induction on the structure of derivations of
  $$\wfiter{\Gamma}{\tau}{\phi?s}{\tau'}$$ and then using inversion and 
  properties of substitutions.
\end{proof}
\begin{theorem}[Completeness]
~
\begin{enumerate}
\item If $\wfupd{\Gamma}{1}{\tau}{\nstmt{s}}{\tau'}$ then
  $\wfupdstmt{\Gamma}{\tau}{s}{\tau'}$.
\item If $\wfupd{\Gamma}{1}{\alpha}{\nupd{u}}{\alpha'}$ then
  $\wfupdsimp{\Gamma}{\alpha}{s}{\alpha'}$.
\item If $\wfupd{\Gamma}{1}{\alpha}{\npath{p}(s)}{\alpha'}$ then there
  exist $\alpha'',\Theta,\Theta'$ such that $\alpha' =
  \alpha''\subst{\Theta}$,
  $\wfupdpath{\Gamma}{\alpha}{p}{\alpha'}{\Theta}$, and
  $\wfupds{\Theta}{s}{\Theta'}$.
\item If $\wfupds{\Theta}{\npath{p}(s)}{\Theta'}$ then there exists
  $\Theta_1,\Theta_2,\Theta_2'$ such that $\Theta' =
  \Theta_1\subst{\Theta_2'}$,
  $\wfupdpaths{\Theta}{p}{\Theta_1}{\Theta_2}$, and
  $\wfupds{\Theta_2}{s}{\Theta_2'}$.
\end{enumerate}
\end{theorem}
\begin{proof}
  Parts (3) and (4) follow by simultaneous induction using previous
  lemmas.  Parts (1) and (2) then follow by simultaneous induction,
  using parts (3) and (4).
\end{proof}

\begin{figure}
  \fbox{$\wfupdstmt{\Gamma}{\tau}{s}{\tau'}$}
  \[\begin{array}{c}
    \infer{\wfupdstmt{\Gamma}{\tau}{s_1;s_2}{\tau''}}{\wfupdstmt{\Gamma}{\tau}{s_1}{\tau'} & \wfupdstmt{\Gamma}{\tau'}{s_2}{\tau''}}
    \smallskip\\
    \infer{\wfupdstmt{\Gamma}{\tau}{\IFTHEN{e}{s}}{\tau|\tau'}}
    {\wf{\Gamma}{e}{\kbool} & \wfupdstmt{\Gamma}{\alpha}{s}{\tau'}}
    \smallskip\\
    \infer{\wfupdstmt{\Gamma}{\tau}{\LETIN{x=e}{s}}{\tau'}}{\wf{\Gamma}{e}{\tau_0} & \wfupdstmt{\Gamma,x:\tau+_0}{\tau}{s}{\tau'}}  
\smallskip\\
\infer{\wfupdstmt{\Gamma}{\alpha}{u}{\alpha'}}{\wfupdsimp{\Gamma}{\alpha}{u}{\alpha'}}
  \end{array}\]  
\caption{Typechecking rules for compound updates}\labelFig{source-tc-stmt}
\end{figure}

\begin{figure}
\fbox{$\wfupdsimp{\Gamma}{\alpha}{u}{\alpha'}$}
  \[\small\begin{array}{c}
    \infer{\wfupdsimp{\Gamma}{\alpha}{\insertvalue{\kbefore}{p}{e}}{\alpha'\subst{\Theta'}}}
{\wfupdpath{\Gamma}{\alpha}{p}{\alpha'}{\Theta} & \wfupds{\Theta}{\kleft[\kinsert~e]}{\Theta'}}
\smallskip\\
    \infer{\wfupdsimp{\Gamma}{\alpha}{\insertvalue{\kafter}{p}{e}}{\alpha'\subst{\Theta'}}}
{\wfupdpath{\Gamma}{\alpha}{p}{\alpha'}{\Theta} & \wfupds{\Theta}{\kright[\kinsert~e]}{\Theta'}}
\smallskip\\
    \infer{\wfupdsimp{\Gamma}{\alpha}{\insertinto{\kfirst}{p}{e}}{\alpha'\subst{\Theta'}}}
{\wfupdpath{\Gamma}{\alpha}{p}{\alpha'}{\Theta} & \wfupds{\Theta}{\kchildren[\kleft[\kinsert~e]]}{\Theta'}}
\smallskip\\
    \infer{\wfupdsimp{\Gamma}{\alpha}{\insertinto{\klast}{p}{e}}{\alpha'\subst{\Theta'}}}
{\wfupdpath{\Gamma}{\alpha}{p}{\alpha'}{\Theta} & 
  \wfupds{\Theta}{\kchildren[\kright[\kinsert~e]]}{\Theta'}}
\smallskip\\
\infer{\wfupdsimp{\Gamma}{\alpha}{\delete{p}}{\alpha'\subst{\Theta'}}}
{\wfupdpath{\Gamma}{\alpha}{p}{\alpha'}{\Theta} & 
  \wfupds{\Theta}{\kdelete}{\Theta'}}
\smallskip\\
\infer{\wfupdsimp{\Gamma}{\alpha}{\deletefrom{p}}{\alpha'\subst{\Theta'}}}
{\wfupdpath{\Gamma}{\alpha}{p}{\alpha'}{\Theta} & 
  \wfupds{\Theta}{\kchildren[\kdelete]}{\Theta'}}
\smallskip\\
\infer{\wfupdsimp{\Gamma}{\alpha}{\rename{p}{n}}{\alpha'\subst{\Theta'}}}
{\wfupdpath{\Gamma}{\alpha}{p}{\alpha'}{\Theta} & 
\wfupds{\Theta}{\krename~n}{\Theta'}}
\smallskip\\
\infer{\wfupdsimp{\Gamma}{\alpha}{\replace{p}{e}}{\alpha'\subst{\Theta'}}}
{\wfupdpath{\Gamma}{\alpha}{p}{\alpha'}{\Theta} & 
  \wfupds{\Theta}{\kdelete;\kinsert~e}{\Theta'}}
\smallskip\\
\infer{\wfupdsimp{\Gamma}{\alpha}{\replacein{p}{e}}{\alpha'\subst{\Theta'}}}
{\wfupdpath{\Gamma}{\alpha}{p}{\alpha'}{\Theta} & 
  \wfupds{\Theta}{\kchildren[\kdelete;\kinsert~e]}{\Theta'}}
\smallskip\\
\infer{\wfupdsimp{\Gamma}{\alpha}{\update{p}{s}}{\alpha'\subst{\Theta'}}}
{\wfupdpath{\Gamma}{\alpha}{p}{\alpha'}{\Theta} & 
\wfupdstmts{\Theta}{s}{\Theta'}}
  \end{array}\]
\caption{Typechecking rules for simple updates}\labelFig{source-tc-upd}
\end{figure}

\begin{figure}
  \fbox{$\wfupdpath{\Gamma}{\alpha}{p}{\alpha'}{\Theta}$}
  \[\begin{array}{c}
    \infer{\wfupdpath{\Gamma}{\alpha}{.}{\alpha}{\emptyset}}{}
\smallskip\\
\infer{\wfupdpath{\Gamma}{\alpha}{p/p'}{\alpha'\subst{\Theta_2}}{\Theta_2'}}
{\wfupdpath{\Gamma}{\alpha}{p}{\alpha'}{\Theta_1}
&
\wfupdpaths{\Theta_1}{p'}{\Theta_2}{\Theta_2'}}
\smallskip\\
\infer{\wfupdpath{\Gamma}{\alpha}{p[e]}{\alpha'\subst{\maybe{\Theta}}}{\Theta}}
{\wfupdpath{\Gamma}{\alpha}{p}{\alpha'}{\Theta} & 
\wfs{\Theta}{e}{\kbool}}
\smallskip\\
\infer{\wfupdpath{\Gamma}{\alpha}{x ~\kas~p}{\alpha'}{\Theta\oplus x}}
{\wfupdpath{\Gamma}{\alpha}{p}{\alpha'}{\Theta}}
\smallskip\\
\infer{\wfupdpath{\Gamma}{n[\tau]}{\phi}{n[\tau']}{\Theta}}
{\wfupdfilt{\Gamma}{\tau}{\phi}{\tau'}{\Theta}}
  \end{array}\]
\caption{Typechecking rules for paths}\labelFig{source-tc-path}
\end{figure}

\begin{figure}
  \fbox{$\wfupdfilt{\Gamma}{\tau}{\phi}{\tau'}{\Theta}$}
\[\begin{array}{c}
  \infer{\wfupdfilt{\Gamma}{\emptyseq}{\phi}{\emptyseq}{\emptyset}}{}
  \smallskip\\
  \infer{\wfupdfilt{\Gamma}{\alpha}{\phi}{\alpha}{\emptyset}}{\alpha \not\subty \phi}
  \smallskip\\
  \infer{\wfupdfilt{\Gamma}{\alpha}{\phi}{Z}{Z \mapsto (\Gamma \tri\alpha)}}{\alpha \subty \phi & \text{$Z$ fresh}}
  \smallskip\\
  \infer{\wfupdfilt{\Gamma}{\tau_1,\tau_2}{\phi}{(\tau_1',\tau_2')}{\Theta_1 \uplus \Theta_2}}
{\wfupdfilt{\Gamma}{\tau_1}{\phi}{\tau_1'}{\Theta_1}
& 
\wfupdfilt{\Gamma}{\tau_2}{\phi}{\tau_2'}{\Theta_2}}
  \smallskip\\
  \infer{\wfupdfilt{\Gamma}{\tau_1|\tau_2}{\phi}{\tau_1'|\tau_2'}{\Theta_1 \uplus \Theta_2}}
{\wfupdfilt{\Gamma}{\tau_1}{\phi}{\tau_1'}{\Theta_1}
& 
\wfupdfilt{\Gamma}{\tau_2}{\phi}{\tau_2'}{\Theta_2}}
  \smallskip\\
  \infer{\wfupdfilt{\Gamma}{\tau_1^*}{\phi}{\tau_2^*}{\Theta}}
{\wfupdfilt{\Gamma}{\tau_1}{\phi}{\tau_2}{\Theta}}
\smallskip\\
\infer{\wfupdfilt{\Gamma}{X}{\phi}{\tau'}{\Theta}}
{\wfupdfilt{\Gamma}{E(X)}{\phi}{\tau'}{\Theta}}
\end{array}\]
  \caption{Typechecking rules for path filters}\labelFig{source-tc-filt}
\end{figure}

\begin{figure}
  \fbox{$\wfs{\Theta}{e}{\tau}$}
  \[\begin{array}{c}
    \infer{\wfs{\emptyset}{e}{\tau}}{}
\quad
    \infer{\wfs{\Theta,Z \mapsto (\Gamma \tri \alpha)}{e}{\tau}}
    {\wfs{\Theta}{e}{\tau} & 
      \wf{\Gamma}{e}{\tau}}
  \end{array}\]
  \fbox{$\wfupds{\Theta}{s}{\Theta'}$}
  \[\begin{array}{c}
    \infer{\wfupds{\emptyset}{s}{\emptyset}}{}
\quad
    \infer{\wfupds{\Theta,Z \mapsto (\Gamma \tri \alpha)}{s}{\Theta',Z \mapsto (\Gamma \tri \tau)}}{\wfupds{\Theta}{s}{\Theta'} &\wfupd{\Gamma}{1}{\alpha}{s}{\tau}}
  \end{array}\]
  \fbox{$\wfupdstmts{\Theta}{s}{\Theta'}$}
  \[\small\begin{array}{c}
    \infer{\wfupdstmts{\emptyset}{s}{\emptyset}}{}
\quad
    \infer{\wfupdstmts{\Theta,Z \mapsto (\Gamma \tri \tau)}{s}{\Theta',Z \mapsto (\Gamma \tri \tau')}}{\wfupdstmts{\Theta}{s}{\Theta'} & \wfupdstmt{\Gamma}{\tau}{s}{\tau'}}
  \end{array}\] 
  \fbox{$\wfupdpaths{\Theta}{p}{\Theta'}{\Theta''}$}
  \[\begin{array}{c}
    \infer{\wfupdpaths{\emptyset}{p}{\emptyset}{\emptyset}}{}
\smallskip\\
    \infer{\wfupdpaths{\Theta,Z \mapsto (\Gamma \tri \alpha)}{p}{\Theta',Z \mapsto (\Gamma \tri \alpha')}{\Theta'' \uplus \Theta'''}}
    {\wfupdpaths{\Theta}{p}{\Theta'}{\Theta''} &
      \wfupdpath{\Gamma}{\alpha}{p}{\alpha'}{\Theta'''}}
  \end{array}\] 
 \caption{Simultaneous  typechecking judgments}\labelFig{simult-source-tc}
\end{figure}

\end{document}